\documentclass[aps,
		prd,
		reprint,
        superscriptaddress,
        shortbibliography,
		  nofootinbib,
		floatfix,
		notitlepage
		]{revtex4-2}

\usepackage{fullpage}
\usepackage{bm}
\usepackage{graphicx,color}
\usepackage{xcolor}
\usepackage{physics}
\usepackage{dsfont}
\usepackage[normalem]{ulem}
\usepackage{hyperref,url}
\usepackage{mathrsfs}
\usepackage[scr=boondox,  
            cal=esstix]   
           {mathalpha}
\usepackage{calrsfs}
\usepackage{bbold}
\usepackage{mathtools}
\usepackage{comment}
\usepackage{scalerel}
\usepackage{tikz}

\newcommand{\mathbfs}[1]{\boldsymbol{\mathbf{#1}}}
\newcommand{\jj}{{\mathscr{j}}}

\usepackage{subfig}

\DeclareMathOperator{\diag}{diag}

\renewcommand{\vec}[1]{\mathbf{#1}}
\newcommand{\cyl}{%
\tikz{
  \draw (0,0) ellipse (0.1 and 0.035);
  \draw (-0.1,0) -- (-0.1,-0.2);
  \draw (0.1,0) -- (0.1,-0.2);
  \draw (-0.1,-0.2) arc (180:360:0.1 and 0.035);
  \draw[dashed] (-0.1,-0.2) arc (180:0:0.1 and 0.035);
}}

\usepackage{varioref}

\usepackage[active]{srcltx}

\expandafter\ifx\csname package@font\endcsname\relax\else
\expandafter\expandafter
\expandafter\usepackage
\expandafter\expandafter
\expandafter{\csname package@font\endcsname}%
\fi
\hypersetup{
	colorlinks=true,       
	linkcolor=blue,          
	citecolor=blue,        
}

\usepackage[section]{placeins}
\usepackage{enumitem}
\usepackage{fancybox}
\labelformat{equation}{Eq.~(#1)} 
\labelformat{figure}{Fig.~#1}
\labelformat{section}{Sec.~#1}
\labelformat{appendix}{App.~#1}

\DeclareMathAlphabet{\mathcal}{OMS}{cmsy}{m}{n}
\newcommand{\be}{\begin{equation}}
\newcommand{\ee}{\end{equation}}
\newcommand{\bi}{\begin{enumerate}}
\newcommand{\ei}{\end{enumerate}}

\begin{document}
\title{Entanglement scaling and dynamics in the Sauter-Schwinger effect}
\author{S. Mahesh Chandran} 
\email{maheshchandran@snu.ac.kr}
\affiliation{Seoul National University, Department of Physics and Astronomy, Center for Theoretical Physics, Seoul 08826, Korea}
\author{Karthik Rajeev}
\email{karthik.rajeev@ed.ac.uk}
\affiliation{Higgs Centre, School of Physics and Astronomy, University of Edinburgh, EH9 3FD, UK}

\begin{abstract}
In quantum field theory, entanglement entropy under spatial bipartitioning serves as a powerful information-theoretic probe of quantum correlations. 
In this work, we present the first comprehensive numerical study of the dynamical evolution and geometric scaling of entanglement entropy in a nonperturbative, strong-field QED setting --- specifically, in the context of the Sauter–Schwinger effect. While the weak-field regime is dominated by area-law states, we show that the entanglement entropy undergoes a transition from area-law to a volume-law scaling for certain strong-field regimes in the pulse-profile parameter space --- signaling a 
fundamental shift in the underlying correlation structure induced by nonperturbative pair production. For intermediate regimes, the scaling is a power-law that interpolates between area- and volume-law behavior. Finally, we provide interpretations based on the behavior of the low-energy pair-creation spectrum and discuss how these insights could inform future investigations of related phenomena.
\end{abstract}

\maketitle

\section{Introduction}

Entanglement pertains to the intrinsically quantum correlations between subsystems, and provides a powerful framework for probing the foundational aspects of a wide range of physical phenomena~\cite{2008Amico.etalRMP,2009Horodecki.etalRMP} --- from microscopic quantum computing
~\cite{2010Nielsen.Chuang}, to black hole thermodynamics~\cite{1986Bombelli.etalPRD,1993SrednickiPRL,2008Das.etal,2011SolodukhinLRiR,2020Chandran.ShankaranarayananPRD} and the dynamics of cosmological perturbations in the early Universe~\cite{2008Campo.ParentaniPRD,2013Maldacena.PimentelJoHEP,2020Brahma.etalPRD,2024Chandran.etalPRD,2025Belfiglio.etalPR}. In recent years, there has been growing interest in employing entanglement as a diagnostic tool to uncover quantum effects in collider physics and related phenomenological contexts (for a review, see \cite{2024Barr.etalPiPaNP}). Similarly, in quantum electrodynamics (QED), both in vacuum and under strong-field conditions, systematic studies of entanglement generated in physical processes are increasingly being explored~\cite{2017CerveraLierta.etalSP,2023Fedida.SerafiniPRD,2024Fedida.etalPRD,2025Tang.etalPRD}. Despite this progress, the role of entanglement in the nonperturbative regimes of local quantum field theories remains 
relatively unexplored, despite its fundamental significance. In this work, we focus on one of the most studied nonperturbative phenomena in QED --- the Sauter–Schwinger effect~\cite{1931SauterZfP,1936Heisenberg.EulerZfP,1951SchwingerPR} --- and investigate its implications for the quantum correlations of the decaying vacuum.

Recent advances in high-intensity laser technology are creating new opportunities to probe nonperturbative quantum phenomena in extreme electromagnetic fields. Next-generation ultra-high-intensity lasers are expected to approach the critical Schwinger field strength, opening unprecedented possibilities to study strong-field QED effects, including vacuum pair production~\cite{2016Turcu.etalRRP,2021Yujie.etalTRoLE}. In parallel, condensed-matter analogues have been shown to provide tunable, scalable platforms for modeling strong-field QED effects~\cite{2012Szpak.SchuetzholdNJoP,2023Schmitt.etalNP,2024Adorno.etalPRB}. These experimental prospects highlight the urgency of theoretical studies
characterizing the quantum properties of the vacuum under such extreme conditions.

Among various approaches that have proven useful for quantifying entanglement
, the entanglement entropy under spatial bipartition stands out as a particularly powerful tool for characterizing the quantum correlations of vacuum and excited states in quantum field theory~(QFT). For local dynamics, it is well known that the leading contribution to the entanglement entropy of vacuum states typically scales with the area of the boundary separating the subsystems. Historically, this area law behavior was first identified in QFT~\cite{1986Bombelli.etalPRD,1993SrednickiPRL,2004Calabrese.CardyJoSMTaE,2009Casini.HuertaJoPAMaT} and later found to hold broadly across gapped systems~\cite{2010Eisert.etalRMP}, with various subleading corrections arising from geometry~\cite{2008SolodukhinPLB,2020Benedetti.CasiniPRD}, topology~\cite{2006Kitaev.PreskillPRL,2006Levin.WenPRL}, higher field-derivatives~\cite{2017Kumar.ShankaranarayananSR} and
cosmological long-range correlations~\cite{2013Maldacena.PimentelJoHEP}. In contrast, a “typical” pure state in Hilbert space exhibits a volume-law scaling of entanglement entropy under spatial bipartition --- a direct consequence of Page’s theorem \cite{1993PagePRL}. This distinction implies that vacuum states are, in this precise sense, highly atypical among all possible pure quantum states.

While typicality is often formalized by averaging the entanglement entropy over all possible quantum states accessible under unitary evolution~\cite{1993PagePRL,2022Bianchi.etalPQ}, the exact dynamics can depend crucially on the specifics of the physical process in question. Entanglement entropy in bosonic systems --- studied via controlled quenches~\cite{2016Cotler.etalJoHEP,2017Caputa.etalPLB,2018Alba.CalabreseSP,2018Hackl.etalPRA} and mode squeezing~\cite{2015Bianchi.etalPRD,2024Katsinis.etalJoHEP} --- has to this end provided the tools to address various phenomenological aspects pertaining to the dynamical Casimir effect~\cite{2019Romualdo.etalPRD,2023Chandran.ShankaranarayananPRD,2025Good.etalPoTaEP}, cosmological perturbations~\cite{2024Chandran.etalPRD,2024Boutivas.etalJoCaAP,2025AndrzejewskiJoHEP} and gravitational collapse~\cite{2025Belfiglio.etalPRDa}. Despite these advances, the relationship between nonperturbative pair creation in strong fields --- a fundamental prediction of QFT --- and spatial entanglement remains
poorly understood, with a systematic treatment still lacking. 

In this work, we present the first comprehensive numerical investigation of the dynamical evolution
and geometric scaling of entanglement entropy in a nonperturbative strong-field QED setting --- specifically, in the context of the Sauter–Schwinger effect. To this end, we consider the evolution of the vacuum state of scalar QED in the presence of homogeneous Sauter pulse of electric field, and 
track the entanglement entropy of cylindrical subregions carved out by a spatial bipartition.  We demonstrate that, for sufficiently strong background fields and large regions of the parameter space characterizing the pulse-profile, the entanglement entropy transitions from an area-law to volume-law scaling, signaling a
fundamental shift in the underlying quantum correlation structure indicative of typicality. 

The paper is organized as follows --- In \ref{sec:warmup}, we outline the basic quantum information–theoretic framework employed in our analysis, followed in \ref{sec:momentumspace_detailed} by a brief review of the Sauter–Schwinger effect for a complex scalar field in a homogeneous Sauter pulse, including a mode-by-mode analysis of pairwise entanglement. In \ref{sec:realspace_detailed}, we develop the formalism to investigate the geometric entanglement generated by nonperturbative pair production in this background and present our key results. Finally, \ref{sec:discussion} concludes with a summary and an outlook for future work.

\section{Primer on Gaussian states and entanglement}\label{sec:warmup}

The dynamics of scalar-QED vacuum under the influence of a time-dependent electric field background can be fully described in terms of Gaussian states. The specific details of this description will be outlined in \ref{sec:momentumspace_detailed}. In this section, therefore, we provide a quick overview of some Gaussian-state essentials, including the basics of entanglement entropy 
under a bipartition of modes.

\subsection{Multimode bosonic Gaussian states}
A multimode bosonic system is associated with a tensor product Hilbert-space $\mathcal{H}=\mathcal{H}_{\rm b}^{\otimes N}$, where $\mathcal{H}_{\rm b}$ is the Hilbert-space for a single bosonic oscillator. The system can be described by the quadrature field operators $\{\hat{\varphi}_i,\hat{\pi}_i\}^{N}_{i=1}$ that satisfy $[\hat{\varphi}_{i},\hat{\pi}_j]=i\delta_{ij}$ and $[\hat{\varphi}_{i},\hat{\varphi}_j]=[\hat{\pi}_{i},\hat{\pi}_j]=0$. It is convenient to collect these operators into a $2N-$dimensional vector $\hat{\mathbf{Q}}$, with components $\hat{Q}_i=\hat{\varphi}_i$ and $\hat{Q}_{i+N}=\hat{\pi}_i$.
The commutation relations for the quadrature field operators can then be represented concisely as follows:
\begin{equation}
    [\hat{Q}_i,\hat{Q}_j]=i\Omega_{ij}\,,\quad \mathbf{\Omega}=\begin{bmatrix}
        O&\mathbb{I}\\-\mathbb{I}&O
    \end{bmatrix}\,,
\end{equation}
where $\mathbf{\Omega}$ is the symplectic matrix and $\mathbb{I}$ is the $N\times N$ identity matrix.

For any density matrix $\hat{\rho}$, one can define the Wigner characteristic function~\cite{2005Braunstein.LoockRMP}
\begin{align}
    \mathcal{W}(\mathbf{q})={\rm Tr}\!\!\left[\!\hat{\rho}\,\exp\!\!\left(\!i\sum_{i,j=1}^{2N}\hat{Q}_i\,\Omega_{ij}\,q_{j}\!\right) \!\!\right]\,\,\,;\,\,\,\mathbf{q}\in\mathbb{R}^{2N},
\end{align}
which provides an equivalent phase-space characterization of the quantum state described by $\hat{\rho}$. Gaussian states are defined as those whose Wigner characteristic function $\chi(\mathbf{q})$ is Gaussian. For such states, the mean $\bar{\mathbf{Q}}={\rm Tr}[\mathbf{\hat{Q}}\hat{\rho}]$ and the covariance matrix $\mathbf{\Sigma}$, defined as~\cite{2010Eisert.etalRMP}:
\begin{equation}\label{eq:cov}
    \Sigma_{ij}=\frac{1}{2}{\rm Tr}\left[\{\hat{Q}_i-\bar{Q}_i,\hat{Q}_j-\bar{Q}_j\}\hat{\rho}\right]
\end{equation}
contain the full information about the state. 

It turns out that for the system we are about to consider in the paper $\bar{\mathbf{Q}}=\mathbf{0}$, so we shall be assuming that from here onward. This special class of \textit{zero-mean} Gaussian states are fully characterized by just the covariance matrix $\mathbf{\Sigma}$, which can be conveniently organized into~\cite{2010Eisert.etalRMP}:
\begin{equation}\label{eq:cov1}
    \mathbfs{\Sigma}=\begin{bmatrix}       \mathbfs{\Sigma}_{\varphi\varphi}&\mathbfs{\Sigma}_{\varphi\pi}\\\mathbfs{\Sigma}_{\varphi\pi}^T&\mathbfs{\Sigma}_{\pi\pi}    \end{bmatrix}\,,
\end{equation}
where the block matrices in $\mathbf{\Sigma}$ are constructed from two-point correlation functions via
\begin{subequations}\label{eq:corr}
    \begin{align}
        (\Sigma_{\varphi\varphi})_{ij}&=\frac{1}{2}\langle \{\hat{\varphi}_i,\hat \varphi_j\}\rangle
        \\
        (\Sigma_{\varphi\pi})_{ij}&=\frac{1}{2}\langle \{\hat \varphi_i,\hat\pi_j\}\rangle
        \\
        (\Sigma_{\pi\pi})_{ij}&=\frac{1}{2}\langle \{\hat\pi_i,\hat\pi_j\}\rangle,
    \end{align}
\end{subequations}
where we use the notation $\left\langle...\right\rangle={\rm Tr} [...\hat{\rho}]$ and we have used $\bar{\mathbf{Q}}=0$.

\subsection{Entanglement entropy}\label{subsec:realspace_review}
For an arbitrary multimode \textit{zero-mean} Gaussian state, the different modes of the system would generally be entangled. One approach to probe these inter-mode quantum correlations is to collect a set of modes, say $\mathcal{D}_{in}=\{i|i\leq n<N\}$, to define an IN-subsystem. The idea is then to investigate the entanglement between the IN-subsystem and the rest of the system (OUT-subsystem).

The reduced density matrix $\hat{\rho}_{in}$ of IN-subsystem will also be \textit{zero-mean} Gaussian, characterized by the reduced covariance matrix $\mathbfs{\sigma}$, obtained simply from \ref{eq:cov} by taking the $n\cross n$ sub-blocks of the $N\cross N$ matrices in \ref{eq:corr}, such that
\begin{subequations}\label{eq:def_sigma_in}
    \begin{align}
\mathbfs{\sigma}&=\begin{bmatrix}       \mathbfs{\sigma}_{\varphi\varphi}&\mathbfs{\sigma}_{\varphi\pi}\\\mathbfs{\sigma}_{\varphi\pi}^T&\mathbfs{\sigma}_{\pi\pi}    \end{bmatrix}\,,\\
\left(\sigma_{\varphi\varphi},\sigma_{\pi\pi},\sigma_{\varphi\pi}\right)_{ij}&=\left(\Sigma_{\varphi\varphi},\Sigma_{\pi\pi},\Sigma_{\varphi\pi}\right)_{ij\in\mathcal{D}_{in}}\,.
    \end{align}
\end{subequations}
The entanglement entropy $S$, which measures the entanglement between the IN- and the OUT-subsystems, is then defined as
\begin{align}\label{eq:S_in}
    S=-{\rm Tr}\left[\hat{\rho}_{in}\log(\hat{\rho}_{in})\right]\,.
\end{align}
The entanglement entropy can be computed from the symplectic eigenvalues of $\mathbfs{\sigma}$, defined as the eigenvalues of the matrix $i\mathbfs{\Omega}_{in}\mathbfs{\sigma}$, where $\mathbfs{\Omega}_{in}$ is the symplectic matrix associated with the IN-subsystem. Specifically, consequent to Williamson's theorem~\cite{1936WilliamsonAJoM}, there exists a symplectic transformation $\mathbf{M}$ such that
\begin{equation}\label{eq:tilde_sigma_in}
    \tilde{\mathbfs{\sigma}}\!=\!\mathbf{M}\mathbfs{\sigma} \mathbf{M}^T\!=\!\begin{bmatrix}
        \textrm{diag}(\gamma_1,..,\gamma_n)&O\\O&\textrm{diag}(\gamma_1,..,\gamma_n)
    \end{bmatrix},
\end{equation}
where ${\rm diag}(...)$ denotes a diagonal matrix and $\{\gamma_k\}$ is the symplectic spectrum, obtained from the eigenvalues $\{\pm \gamma_k\}$ of the matrix $i\mathbfs{\Omega}_{in}\mathbfs{\sigma}$. Thus, the entanglement entropy finally reads
\begin{multline}\label{eq:Entdef}
    S=\sum_{k=1}^n\Bigg[\left(\gamma_k+\frac{1}{2}\right)\log\left(\gamma_k+\frac{1}{2}\right)-\\\left(\gamma_k-\frac{1}{2}\right)\log\left(\gamma_k-\frac{1}{2}\right)\Bigg]\,.
\end{multline}

So far, our discussion has focused on systems with a finite number of bosonic modes. For applications to quantum field theory (QFT), the framework must be judiciously extended to encompass an infinite number of modes, i.e., \( N \rightarrow \infty \). In this continuum limit, modulo well-known caveats~\cite{1996Haag}, the discrete mode indices \((i, j)\) are naturally replaced by continuous field modes of the scalar field, characterized by the momentum vector (or, equivalently, by a set of angular momentum quantum numbers together with the radial energy). In the following section, we review the Sauter–Schwinger effect for a complex scalar field subject to a homogeneous Sauter pulse, working in a basis adapted to the cylindrical symmetry of the problem, and identify the natural mode labeling prescription that enables the application of the formalism outlined above.

\section{Sauter-Schwinger effect and pairwise entanglement}\label{sec:momentumspace_detailed}

We start by writing the Minkowski metric in the cylindrical coordinates as
\begin{align}
    ds^2=dt^2-dz^2-d\varrho^2-\varrho^2d\phi\,.
\end{align}
The Sauter-pulse background can then be described by the gauge potential
\begin{align}\label{eq:def_sauter_A}
    A(t)=-E_0\tau\left[1+\tanh{\left(\frac{t}{\tau}\right)}\right]dz
\end{align}
which gives rise to the electromagnetic field
\begin{align}\label{eq:def_sauter_F}
    F(t)= \frac{E_0}{\cosh^2\left(t/\tau\right)}dz\wedge dt
\end{align}
that describes a homogeneous electric field of magnitude $E_0$ along the $z-$direction, and a temporal pulse width of the order $\tau$.

Before we move on to reviewing pair creation of a complex scalar field in this background, it is useful to briefly recall the dimensionless parameters that characterize the system. Denoting the mass and charge of the scalar particle by $\mu$ and $e$, respectively, these parameters may be chosen as follows~\cite{2023Fedotov.etalPR}:
\begin{itemize}
    \item \textbf{Classical nonlinearity (intensity) parameter} $\xi$:
    \begin{equation}
        \xi \equiv \frac{|eE_0|}{\mu\omega}
        = \frac{|eE_0|\tau}{\mu} \, .
    \end{equation}
    The parameter $\xi$ measures the work done by the electric field over a Compton wavelength in units of the characteristic field quantum energy. 
    \item \textbf{Adiabaticity parameter} $\eta$:
    \begin{equation}
        \eta \equiv \frac{\omega}{\mu}
        = \frac{1}{\mu\tau} \, .
    \end{equation}
    This parameter quantifies how rapidly the external field varies compared to the Compton time. The adiabatic regime $\eta\ll1$ corresponds to slowly varying backgrounds, whereas $\eta\gtrsim1$ describes rapidly varying, nonadiabatic fields.
    \item \textbf{Quantum nonlinearity parameter} $\chi$:
    \begin{equation}
        \chi = \frac{|eE_0|}{\mu^2} = \xi\,\eta \, .
    \end{equation}
    In the adiabatic regime, the tunneling probability behaves as $P\propto e^{-\pi/\chi}$, being exponentially suppressed for $\chi\ll1$ and unsuppressed for $\chi\gtrsim1$.
\end{itemize}

\subsection{Mode functions and the asymptotic state}\label{subsec:modes_continuum}

The Klein-Gordon equation for a complex scalar in an external electromagnetic background $A_{\mu}$ takes the form
\begin{align}
    \left[\left(i\partial_{\mu}-eA_{\mu}\right)\left(i\partial^{\mu}-eA^{\mu}\right)-\mu^2\right]\Phi=0\,.
\end{align}
For the Sauter-pulse background in \ref{eq:def_sauter_A}, it is customary to choose a basis of solutions to the above equation parameterized by the spatial momentum $\mathbf{k}$. However, when studying the geometrical entanglement entropy, the symmetry of the system makes a cylindrical mode basis more appropriate. 

To make the notation closer to that of the standard plane-wave basis, we shall use the short-hand $\mathbf{k}=(n,l,m)$ and $\tilde{\mathbf{k}}\equiv(n,-l,-m)$, so that the positive/negative energy cylindrical modes, appropriate to asymptotic past, can be written as 
\begin{subequations}
    \begin{align}
    \mathcal{P}^{\pm}_{\mathbf{k}}(x)&=\mathcal{R}_{\mathbfs{k}}(\mathbf{x})v^{\pm}_{\mathbf{k}}(t)\,,\\
    \mathcal{R}_{\mathbfs{k}}(\mathbfs{x})&\equiv\mathcal{N}_{\mathbf{k}}e^{i\frac{2\pi l}{L}z}e^{im\phi}J_{m}\!\left(\!\frac{\jj_{mn}\varrho}{R}\!\right)\,,
\end{align}
\end{subequations}
where $m,l\in\mathbb{Z}$, $n\in\mathbb{Z}^+$, $J_{m}(x)$ is the Bessel function, $\jj_{mn}$ denotes the $n^{\rm th}$ root of $J_{m}(x)$, the normalization $\mathcal{N}_{\mathbf{k}}=2\sqrt{\pi}/|J_{m+1}(\jj_{mn})|$ and $v^{\pm}_{\mathbf{k}}(t)$ satisfies the time-dependent harmonic oscillator equation
\begin{align}\label{eq:EOM_for_v}
    \ddot{v}^{\pm}_{\mathbf{k}}(t)+\omega_{\mathbf{k}}^2(t)\,v^{\pm}_{\mathbf{k}}(t)=0\,,
\end{align}
with the frequency given by
\begin{align}\label{eq:normal_mode_continuum}
    \omega_{\mathbf{k}}^2(t)=\kappa_{nm}^2+\mu^2+\left(\frac{2\pi l}{L}+eA_{z}(t)\right)^2\,.
\end{align}
where, $\kappa_{nm}=\jj_{mn}/R$. For the Sauter pulse, the modes functions appropriate to the asymptotic past --- satisfying the boundary condition $v^{\pm}_{\mathbf{k}}(t\sim-\infty)\sim e^{\mp i\omega_{\mathbf{k}}(-\infty)t}$ --- can be solved for exactly and are given by $v^{+}_{\mathbf{k}}(t)=v_{\mathbf{k}}(t)$ and $v^{-}_{\mathbf{k}}(t)=v^*_{\mathbf{k}}(t)$, with~\cite{2008Kim.etalPRD}:
\begin{equation}\label{eq:modefn}
    v_{\mathbf{k}}=(1+e^{\frac{2t}{\tau}})^{\frac{1}{2}+i\lambda}{}_{2}F_{1}\left(\gamma_{\mathbfs{k}}^{-},\gamma_{\mathbfs{k}}^{+};\gamma_{\mathbfs{k}};-e^{\frac{2t}{\tau}}\right)\frac{e^{-i\omega^{-}_{\mathbfs{k}}t}}{\sqrt{2\omega^{-}_{\mathbfs{k}}}},
\end{equation}
where we define:
\begin{subequations}
\begin{align}
    \omega^{\pm}_{\mathbfs{k}}&=\lim_{t\rightarrow\pm\infty}\omega_{\mathbfs{k}}(t)\\
    \gamma_{\mathbfs{k}}^{\pm}&=\frac{1}{2}-\frac{i}{2}\left(\omega^{-}_{\mathbfs{k}}\pm\omega^{+}_{\mathbfs{k}}\right)\tau+i\lambda\\
    \gamma_{\mathbfs{k}}&=1-i\omega^{-}_{\mathbfs{k}} \tau\\
    \lambda&=\sqrt{\frac{\xi^2}{\eta^2}-\frac{1}{4}}\,.
\end{align}
\end{subequations}
The modes satisfy appropriate completeness and orthonormality relations as described in Appendix~\hyperref[app:cylmodes_ortho]{A}.

The complex scalar field operator can be expanded in the above basis to get
\begin{subequations}
\begin{align}
    \hat{\Phi}(x)&=\sum_{\mathbf{k}}\left[b_{\mathbf{k}}\mathcal{P}^+_{\mathbf{k}}(x)+d^{\dagger}_{\tilde{\mathbf{k}}}\mathcal{P}^-_{\mathbf{k}}(x)\right]\,,\\
       \sum_{\mathbf{k}}&=\frac{1}{\mathcal{V}}\sum_{m=-\infty}^{\infty}\sum_{l=-\infty}^{\infty}\sum_{n=0}^{\infty}\,,
\end{align}  
\end{subequations}
where $\mathcal{V}=\pi R^2 L$ is the volume of the system.  The creation/annihilation operators satisfy
\begin{align}\label{eq:commut_rel}
    [b_{\mathbfs{k}},b^{\dagger}_{\mathbfs{k}'}]&=[d_{\mathbfs{k}},d^{\dagger}_{\mathbfs{k}'}]=\mathcal{V}\delta_{mm'}\delta_{ll'}\delta_{nn'}\equiv \mathcal{V}\delta_{\mathbf{k}\mathbf{k}'}
\end{align}
with all the other commutators vanishing. One then defines the $in-$vacuum $\ket{\rm in}$ as the state annihilated by $\{b_{\mathbfs{k}},d_{\mathbfs{k}}\}$, for all choices of $\mathbfs{k}$.
\begin{align}
   b_{\mathbfs{k}}\ket{\rm in}=d_{\mathbfs{k}}\ket{\rm in}=0\quad,\quad\forall \mathbfs{k}\,. 
\end{align}
Having established the basic framework for quantizing a complex scalar field in a Sauter pulse background, we now turn to the discussion of pair creation and related aspects.

\subsection{Particle Number}

The scalar field can equivalently be expanded in terms of the asymptotic future modes, $\mathcal{F}^{\pm}_{\mathbf{k}}(x)$, given by
\begin{align}
    \mathcal{F}^{\pm}_{\mathbf{k}}(x)=\mathcal{R}_{\mathbfs{k}}(\mathbfs{x})\,u^{\pm}_{\mathbf{k}}(t)\,,
\end{align}
where $u^{\pm}_{\mathbf{k}}(t)$ also solves \ref{eq:EOM_for_v} but with the boundary condition $u^{\pm}_{\mathbf{k}}(t\sim \infty)\sim e^{\mp i\omega^{+}_{\mathbf{k}}t}$. The scalar field operator then takes the equivalent form
\begin{align}
    \hat{\Phi}(x)&=\sum_{\mathbf{k}}\left[B_{\mathbf{k}}\mathcal{F}^+_{\mathbf{k}}(x)+D^{\dagger}_{\tilde{\mathbf{k}}}\mathcal{F}^-_{\mathbf{k}}(x)\right]
\end{align}
with the set of operators $\{B_{\mathbf{k}},B^{\dagger}_{\mathbf{k}},D_{\mathbf{k}},D^{\dagger}_{\mathbf{k}}\}$ satisfying the same commutation relations as in \ref{eq:commut_rel}. The $out-$vacuum $\ket{\rm out}$ as the state annihilated by $\{B_{\mathbfs{k}},D_{\mathbfs{k}}\}$, for all choices of $\mathbfs{k}$.
\begin{align}
   B_{\mathbfs{k}}\ket{\rm out}=D_{\mathbfs{k}}\ket{\rm in}=0\quad,\quad\forall \mathbfs{k}\,. 
\end{align}

Since both the past modes ($\mathcal{P}^{\pm}_{\mathbf{k}}$) and future modes ($\mathcal{F}^{\pm}_{\mathbf{k}}$) independently form a complete basis, they are linearly related, such that
\begin{subequations}
  \begin{align} \mathcal{P}^{+}_{\mathbf{k}}&=\alpha_{\mathbf{k}}\mathcal{F}^{+}_{\mathbf{k}}+\beta_{\mathbf{k}}^*\mathcal{F}^{-}_{\mathbf{k}}\\
\mathcal{P}^{-}_{\mathbf{k}}&=\alpha_{\mathbf{k}}^*\mathcal{F}^{-}_{\mathbf{k}}+\beta_{\mathbf{k}}\mathcal{F}^{+}_{\mathbf{k}}\,,    
\end{align}  
\end{subequations}
where
\begin{subequations}
\begin{align}
 \alpha_{\mathbf{k}}&=\frac{\Gamma(-i\omega^{+}_{\mathbf{k}}\tau)\Gamma(1-i\omega^{-}_{\mathbf{k}}\tau)}{\Gamma(\gamma^{-}_{\mathbf{k}})\Gamma(\gamma_{\mathbf{k}}-\gamma^{+}_{\mathbf{k}})}\sqrt{\frac{\omega^{+}_{\mathbf{k}}}{\omega^{-}_{\mathbf{k}}}}\,,\\ \beta_{\mathbf{k}}&=\frac{\Gamma(-i\omega^{+}_{\mathbf{k}}\tau)\Gamma(1+i\omega^{-}_{\mathbf{k}}\tau)}{\Gamma(\gamma^{+*}_{\mathbf{k}})\Gamma(\gamma^*_{\mathbf{k}}-\gamma^{-*}_{\mathbf{k}})}\sqrt{\frac{\omega^{+}_{\mathbf{k}}}{\omega^{-}_{\mathbf{k}}}}\,.
\end{align}
\end{subequations}
The creation/annihilation operators are, in turn, related via
\begin{align}
    B_{\vec{k}}&=\alpha_{\vec{k}} b_{\vec{k}}+\beta_{\vec{k}}d_{\tilde{\vec{k}}}^\dagger;\quad D_{\tilde{\vec{k}}}^\dagger=\alpha^*_{\vec{k}} d_{\tilde{\vec{k}}}^\dagger+\beta^*_{\vec{k}}b_{\vec{k}}\,.
\end{align}
With the $out-$particle/antiparticle number operators defined, respectively, as
\begin{align}
    N^{+}=\sum_{\mathbfs{k}}B^{\dagger}_{\mathbf{k}}B_{\mathbf{k}}\quad;\quad N^{-}=\sum_{\mathbfs{k}}D^{\dagger}_{\mathbf{k}}D_{\mathbf{k}}\,.
\end{align}
The average number of particles produced is then given by
\begin{align}
   \langle{\rm in}|N^{\pm}|{\rm in}\rangle=\mathcal{V}\sum_{\mathbf{k}}|\beta_{\mathbf{k}}|^2\equiv \mathcal{V}\sum_{\mathbf{k}}n^{\pm}_{\mathbf{k}}
\end{align}
For the Sauter pulse, the spectrum $n_{\mathbf{k}}$ takes the form
\begin{equation}
    n^{+}_{\mathbf{k}}=n^{-}_{\mathbf{\tilde{k}}}=\frac{\cosh{[\pi(\omega^{-}_{\mathbf{k}}-\omega^{+}_{\mathbf{k}} )\tau]}+\cosh[2\pi\lambda]}{2\sinh[\pi\omega^{-}_{\mathbf{k}}\tau]\sinh[\pi\omega^{+}_{\mathbf{k}} \tau]}\,.
\end{equation}
It is well known that in the adiabatic regime ($\eta\rightarrow0$) and the limit of small transverse momentum for the outgoing particles the above spectrum reduces to that of the Schwinger effect~\cite{1996Gavrilov.GitmanPRD}:
\begin{equation}
    n_\vec{k}^+\sim \exp\left[-\frac{\pi}{\chi}\left(1+\frac{\kappa^2}{\mu^2}\right)\right]\,.
\end{equation}
Several works have examined possible thermal interpretations of the above spectrum. For example, in the case $\kappa=0$, it can be viewed as the Boltzmann factor of particles with energy $\mu$ and temperature $\beta^{-1}=eE/\pi \mu$~\cite{1981RitusSPJ,1977Mueller.etalPLA,1996Gavrilov.GitmanPRD}. Evidently, this does not apply generally for $\kappa \neq 0$.

\begin{figure}[!htb]
\vspace*{1em}
	\begin{center}
		\centering
	\includegraphics[scale=0.5]{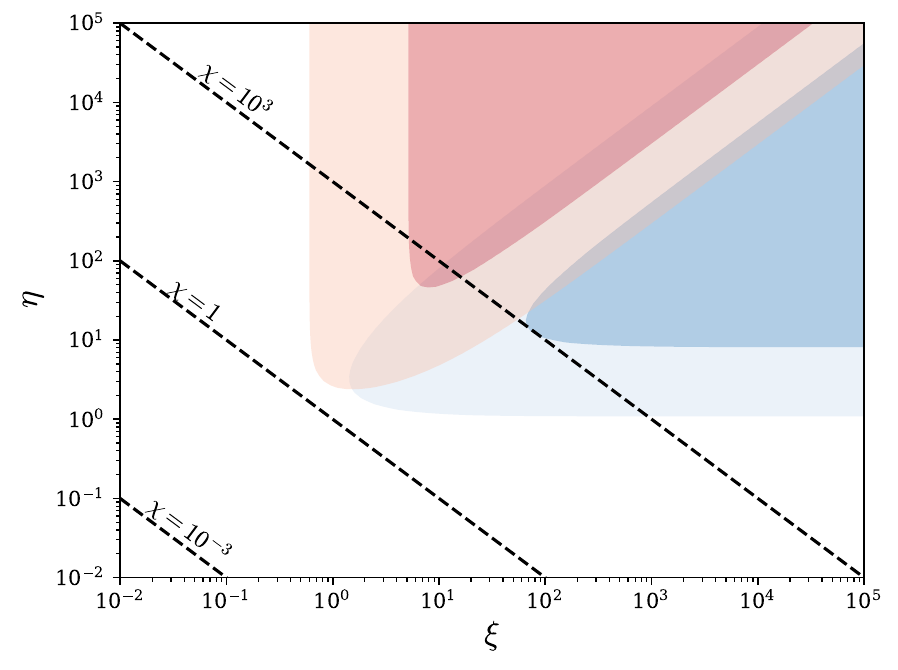}
				\caption{Overlap ($\zeta$) of the Sauter-Schwinger spectrum for the IR mode ($\omega_{\vec{k}}^+=\mu$) with the limiting forms in \ref{eq:regimes} corresponding to Regimes A (red) and B (blue). The contour step size is 0.495, with $\zeta>0.99$ indicating high overlap.}
		\label{fig:thermality}
	\end{center}
\end{figure}

It turns out that for interpreting some of the results presented below, it is useful to consider a different set of limits, focusing on the IR sector ($\omega^{+}_{\bf k}\approx \mu$). For ease of analysis, we note that the transverse \textit{physical} momenta are $p_z^-=k_z+2\mu\xi$ and $p_z^+=k_z$, at asymptotic past and future, respectively. Now, in the strong field $\mu\xi\gg \omega_{\vec{k}}^+$ and nonadiabatic $\mu\eta\gg \omega_{\vec{k}}^+$ regime, the spectrum reduces as follows:
\begin{subequations}\label{eq:thermality}
\begin{align}
    n_{\vec{k}}^+\approx& \frac{1}{\beta\omega_{\vec{k}}^+}\,,\\
    \beta^{-1}=&\frac{\mu\eta}{2\pi}\bigg[\coth{\left(\frac{2\pi \xi}{\eta}\right)}\\+&\cosh{\left(2\pi\sqrt{\frac{\xi^2}{\eta^2}-\frac{1}{4}}\right)}\csch{\left(\frac{2\pi\xi}{\eta}\right)}\bigg].\nonumber   
\end{align}
\end{subequations}
Although the overall spectrum is non-thermal, we see that it coincides with the IR tail of a Planckian distribution in the aforementioned regime. One may interpret this as a near-thermal occupation of the IR modes of the final state corresponding to a temperature $\beta^{-1}$. Depending on how the strong-field limit is taken relative to the nonadiabatic limit, the spectrum further reduces to the following limiting forms:
\begin{equation}\label{eq:regimes}
    n_{\vec{k}}^+\to\begin{cases}
        \frac{\mu\xi}{2\omega_{\vec{k}}^+} &  1\leq \frac{\omega_\vec{k}^+}{\mu}\ll\xi \ll \eta\quad \text{Regime A}\\
        \frac{\mu\eta}{\pi\omega_{\vec{k}}^+} &  1\leq \frac{\omega_\vec{k}^+}{\mu}\ll\eta \ll \xi \quad \text{Regime B}
    \end{cases}
\end{equation}
We observe that in Regime A the spectrum is independent of $\eta$, while in Regime B it is independent of $\xi$ --- implying that upon fixing any one of these parameters and tuning up the other, the particle content in the IR sector asymptotes to a constant value. This also results in an ``L-shaped" symmetry for the low-energy spectrum along the $(\xi,\eta)$ parameter space. To better visualize these regimes, let us consider an ``overlap parameter" $\zeta$ between $n_{\mathbf{k}}(\omega)$ and an arbitrary spectrum $n_X(\omega)$:
\begin{equation}
    \zeta[n_{\mathbf{k}},n_X]=\frac{4n_{\mathbf{k}}n_{X}}{(n_{\mathbf{k}}+n_{X})^2},
\end{equation}
where $\zeta=1$ indicates complete spectral overlap. In \ref{fig:thermality}, we see regions in the $(\xi,\eta)$ parameter space where the IR mode ($\omega_{\vec{k}}^+=\mu$) of the Sauter-Schwinger spectrum overlaps with the limiting forms of \ref{eq:regimes} corresponding to Regimes A and B.

\subsection{Pairwise entanglement}

Let us now define the quadrature operators in the momentum space via
\begin{align}
\begin{pmatrix}
    \hat{\Phi}_{\mathbfs{k}}\\
    \hat{\Pi}_{\mathbfs{k}}
\end{pmatrix}\equiv\int_{\cyl}d^3\mathbf{x}\,\mathcal{R}^*_{\mathbfs{k}}(\mathbfs{x})\begin{pmatrix}
    \hat{\Phi}(x)\\
    \partial_{t}\hat{\Phi}(x)
\end{pmatrix}\,,
\end{align}
where
\begin{align}
   \int_{\cyl}d^3\mathbf{x}(...)=\int_{0}^{R}\varrho d\varrho\int_{0}^{2\pi} d\phi \int_{-L/2}^{L/2}dz(...) \,
\end{align}
is the spatial integral over the finite volume $\mathcal{V}$ of the system. Orthogonality of the set $\{\mathcal{R}_{\mathbfs{k}}(\mathbfs{x})\}$ in this domain leads to 
\begin{subequations}
\begin{align}
    \hat{\Phi}_{\mathbfs{k}}&= b_{\mathbf{k}}v^+_{\mathbf{k}}+d^{\dagger}_{\tilde{\mathbf{k}}}v^-_{\mathbf{k}};\quad \hat{\Phi}_{\mathbfs{k}}^\dagger= b_{\mathbf{k}}^\dagger v^-_{\mathbf{k}}+d_{\tilde{\mathbf{k}}}v^+_{\mathbf{k}}\\
    \hat{\Pi}_{\mathbfs{k}}&=b_{\mathbf{k}}\dot{v}^+_{\mathbf{k}}+d^{\dagger}_{\tilde{\mathbf{k}}}\dot{v}^-_{\mathbf{k}};\quad \hat{\Pi}_{\mathbfs{k}}^\dagger=b_{\mathbf{k}}^\dagger \dot{v}^-_{\mathbf{k}}+d_{\tilde{\mathbf{k}}}\dot{v}^+_{\mathbf{k}}\,\,,
\end{align}   
\end{subequations}
where, we identify the canonical pairs $(\hat{\Phi}_{\mathbfs{k}},\hat{\Pi}_{\mathbfs{k}}^\dagger)$ and $(\hat{\Phi}^\dagger_{\mathbfs{k}},\hat{\Pi}_{\mathbfs{k}})$.
Then, the non-zero two-point correlations can be found to be:
\begin{subequations}
\begin{align}
    \langle{\rm in}|\{\hat{\Phi}_{\mathbf{k}},\hat{\Phi}_{\mathbf{k}'}^\dagger\}|{\rm in}\rangle&=2|v_{\mathbf{k}}|^2\mathcal{V}\delta_{\mathbf{k}\mathbf{k}'}\\
    \langle{\rm in}|\{\hat{\Phi}_{\mathbf{k}},\hat{\Pi}_{\mathbf{k}'}^\dagger\}|{\rm in}\rangle&=2\Re[v_{\mathbf{k}}\dot{v}_{\mathbf{k}}^*]\mathcal{V}\delta_{\mathbf{k}\mathbf{k}'}\\
    \langle{\rm in}|\{\hat{\Pi}_{\mathbf{k}},\hat{\Pi}_{\mathbf{k}'}^\dagger\}|{\rm in}\rangle&=2|\dot{v}_{\mathbf{k}}|^2\mathcal{V}\delta_{\mathbf{k}\mathbf{k}'}
\end{align}
\end{subequations}
The idea is now to make contact with our discussion in \ref{sec:warmup} on the bosonic modes and the operators $\mathbf{Q}$ therein. However, note that $\hat{\Phi}_{\mathbf{k}}$ and $\hat{\Pi}_{\mathbf{k}}$ are not Hermitian and cannot be used as such to define a covariance matrix basis $\mathbf{Q}$ for the QFT system. Furthermore, to calculate pairwise entanglement, we must choose a basis where the field variables do not mix ladder operators corresponding to the particle/antiparticle sectors ---
for a clear bipartition of the latter~\cite{2022Martin.etalJoCaAP}. We will now examine a few potential options that serve this purpose, and ultimately present the most 
intuitive choice. 

Let us first construct such a basis with the past-modes and the corresponding operators:
\begin{align}
\sqrt{\mathcal{V}}\,\vec{Q}_{\mathcal{P}}&=\{\hat{{\Phi}}_{b,\mathbfs{k}},\hat{{\Phi}}_{d,\mathbfs{\tilde{k}}},\hat{{\Pi}}_{b,\mathbfs{k}},\hat{{\Pi}}_{d,\mathbfs{\tilde{k}}}\}
\end{align}
where
\begin{align}
    \frac{\hat{{\Phi}}_{b,\mathbfs{k}}}{\sqrt{\omega^{+}_{\mathbfs{k}}}}&=v_{\vec{k}}^+\hat{b}_{\vec{k}}+v_{\vec{k}}^-\hat{b}_{\vec{k}}^{\dagger}\,;\,\sqrt{\omega^{+}_{\mathbfs{k}}}\,\hat{{\Pi}}_{b,\mathbfs{k}}=\dot{v}_{\vec{k}}^+\hat{b}_{\vec{k}}+\dot{v}_{\vec{k}}^-\hat{b}_{\vec{k}}^{\dagger}\nonumber\\
\frac{\hat{{\Phi}}_{d,\mathbfs{\tilde{k}}}}{\sqrt{\omega^{+}_{\mathbfs{k}}}}&=v_{\vec{k}}^+\hat{d}_{\vec{\tilde{k}}}+v_{\vec{k}}^-\hat{d}_{\vec{\tilde{k}}}^{\dagger}\,;\, \sqrt{\omega^{+}_{\mathbfs{k}}}\,\hat{{\Pi}}_{d,\mathbfs{\tilde{k}}}=\dot{v}_{\vec{k}}^+\hat{d}_{\vec{\tilde{k}}}+\dot{v}_{\vec{k}}^-\hat{d}_{\vec{\tilde{k}}}^{\dagger}
\nonumber
\end{align}
Note that the operators have been rescaled to ensure that the covariance matrix elements are dimensionless, as seen below:
\begin{subequations}\label{eq:GHDM1}
\begin{align}
    \mathbfs{\Sigma}_{\mathcal{P}}&=\begin{bmatrix}
       \mathbfs{\Sigma}_{\varphi\varphi} &\mathbfs{\Sigma}_{\varphi\pi}  \\
       \mathbfs{\Sigma}_{\varphi\pi}&\mathbfs{\Sigma}_{\pi\pi}
    \end{bmatrix};\,\,\,\mathbfs{\Sigma}_{\varphi\varphi}=\diag[\omega_{\vec{k}}^{+}|v_{\vec{k}}|^2]\\\mathbfs{\Sigma}_{\varphi\pi}&=\diag[\Re(v_{\vec{k}}\dot{v}_{\vec{k}}^*)]\,;\,\,\mathbfs{\Sigma}_{\pi\pi}=\diag\left[\frac{|\dot{v}_{\vec{k}}|^2}{\omega_{\vec{k}}^{+}}\right]
\end{align}
\end{subequations}
We see that the particle/anti-particle partitions themselves factorize into pure states (i.e., wherein purity $P=\Tr[\hat{\rho}^2]$ remains unity) throughout the evolution, and are identically described by the reduced covariance matrix given below:
\begin{subequations}
\begin{align}
\mathbfs{\Sigma}_{b,\vec{k}}=\mathbfs{\Sigma}_{d,\vec{\tilde{k}}}&=\begin{bmatrix}
       (\Sigma_{\varphi\varphi})_{\vec{k}\vec{k}} &(\Sigma_{\varphi\pi})_{\vec{k}\vec{k}}  \\
       (\Sigma_{\varphi\pi})_{\vec{k}\vec{k}} &(\Sigma_{\pi\pi} )_{\vec{k}\vec{k}}
    \end{bmatrix};\\
    \quad P&=
    \frac{1}{\det{2\mathbfs{\Sigma}_{b,\vec{k}}}}=1
\end{align}
\end{subequations}
Such states correspond to Gaussian and Homogeneous Density Matrices (GHDM) discussed in literature~\cite{2008Campo.ParentaniPRD}, and reveal here that the particle/anti-particle sectors partitioned in terms of in-operators are always unentangled. At late-times ($t/\tau \to \infty$), we see that each partition freely evolves as a single mode squeezed state:
\begin{align}  
(\Sigma_{\varphi\varphi})_{\vec{k}\vec{k}'}&=\frac{\delta_{\vec{k}\vec{k}'}}{2}\left[\cosh{2r_{\vec{k}}}+\sinh{2r_{\vec{k}}}\cos{2\vartheta_{\vec{k}}}\right]\nonumber\\
 (\Sigma_{\varphi\pi})_{\vec{k}\vec{k}'}&=\frac{\delta_{\vec{k}\vec{k}'}}{2}\sinh{2r_{\vec{k}}}\sin{2\vartheta_{\vec{k}}}\\
    (\Sigma_{\pi\pi})_{\vec{k}\vec{k}'}&=\frac{\delta_{\vec{k}\vec{k}'}}{2}\left[\cosh{2r_{\vec{k}}}-\sinh{2r_{\vec{k}}}\cos{2\vartheta_{\vec{k}}}\right],\nonumber
\end{align}
where the squeezing parameter ($r_{\vec{k}}$) and squeezing angle ($\vartheta_{\vec{k}}$) are given by:
\begin{equation}\label{eq:squeeze_para}
    \sinh{r_{\vec{k}}}=\sqrt{n_{\vec{k}}}\,;\quad 2\vartheta_{\vec{k}}=\arg{(\alpha_{\vec{k}}\beta_{\vec{k}}^*)}-2\omega_{\vec{k}}^{+}t\,
\end{equation}
However, for this final state, we can also construct a similar basis in terms of the future-modes as:
\begin{align}\label{eq:BDbasis}
\sqrt{\mathcal{V}}\,\vec{Q}_{\mathcal{F}}&=\{\hat{{\Phi}}_{B,\mathbfs{k}},\hat{{\Phi}}_{D,\mathbfs{\tilde{k}}},\hat{{\Pi}}_{B,\mathbfs{k}},\hat{{\Pi}}_{D,\mathbfs{\tilde{k}}}\}\,,
\end{align}
where
\begin{align}
    \frac{\hat{{\Phi}}_{B,\mathbfs{k}}}{\sqrt{\omega^{+}_{\mathbfs{k}}}}&=u_{\vec{k}}^+\hat{B}_{\vec{k}}+u_{\vec{k}}^-\hat{B}_{\vec{k}}^{\dagger}\,;\,\sqrt{\omega^{+}_{\mathbfs{k}}}\,\hat{{\Pi}}_{B,\mathbfs{k}}=\dot{u}_{\vec{k}}^+\hat{B}_{\vec{k}}+\dot{u}_{\vec{k}}^-\hat{B}_{\vec{k}}^{\dagger}\nonumber\\
\frac{\hat{{\Phi}}_{D,\mathbfs{\tilde{k}}}}{\sqrt{\omega^{+}_{\mathbfs{k}}}}&=u_{\vec{k}}^+\hat{D}_{\vec{\tilde{k}}}+u_{\vec{k}}^-\hat{D}_{\vec{\tilde{k}}}^{\dagger}\,;\, \sqrt{\omega^{+}_{\mathbfs{k}}}\,\hat{{\Pi}}_{D,\mathbfs{\tilde{k}}}=\dot{u}_{\vec{k}}^+\hat{D}_{\vec{\tilde{k}}}+\dot{u}_{\vec{k}}^-\hat{D}_{\vec{\tilde{k}}}^{\dagger}
\nonumber
\end{align}
In this new basis, the covariance matrix takes the characteristic form of a two-mode squeezed state:
\begin{align}
    \mathbfs{\Sigma}_{\mathcal{F}}&=\frac{1}{2}\begin{bmatrix}
        \cosh{2r_{\vec{k}}}& C_{\vec{k}} &0 &\tilde{C}_{\vec{k}} \\
        C_{\vec{k}} & \cosh{2r_{\vec{k}}} &\tilde{C}_{\vec{k}} &0 \\
        0 & \tilde{C}_{\vec{k}} & \cosh{2r_{\vec{k}}} & -C_{\vec{k}} \\
        \tilde{C}_{\vec{k}} & 0 & -C_{\vec{k}} & \cosh{2r_{\vec{k}}}
    \end{bmatrix},\nonumber\\ C_{\vec{k}}&=\sinh{2r_{\vec{k}}}\cos{2\vartheta_{\vec{k}}};\,\,\tilde{C}_{\vec{k}}=\sinh{2r_{\vec{k}}}\sin{2\vartheta_{\vec{k}}}
\end{align}
Here, the particle ($B,\vec{k}$) and anti-particle ($D,\vec{\tilde{k}}$) sectors have been symmetrically partitioned into their respective sub-blocks which are now mixed states. That is, for the particle/anti-particle subsystem, the reduced covariance matrix takes the form:
\begin{subequations}
\begin{align}   \mathbfs{\Sigma}_{B,\vec{k}}&=\mathbfs{\Sigma}_{D,\vec{\tilde{k}}}=\frac{1}{2}\begin{bmatrix}
        \cosh{2r_{\vec{k}}}& 0 \\
        0 & \cosh{2r_{\vec{k}}}
    \end{bmatrix}\\ P&=\sech{2r_{\vec{k}}}=\frac{1}{2n_{\vec{k}}+1}
\end{align}    
\end{subequations}
For a single mode covariance matrix, the symplectic eigenvalue can be obtained from purity as $\nu=(2P)^{-1}$, which here takes the form $\nu=n_{\vec{k}}+\frac{1}{2}$. From this, the entropy of pairwise ($k/\tilde{k}$) entanglement can be arrived at using \ref{eq:Entdef}:
\begin{equation}
    S_{\vec{k}}=S_{\vec{\tilde{k}}}=(n_{\vec{k}}+1)\ln{(n_{\vec{k}}+1)}-n_{\vec{k}}\ln{n_{\vec{k}}},
\end{equation}
which coincides with the expression obtained in \cite{2010Lin.etalPRD}, where it was also noted that the total pairwise entropy $\sum_\vec{k}S_{\vec{k}}$ is double that of a real scalar field. 

\begin{figure}[!htb]
\vspace*{1em}
	\begin{center}
		\centering
	\includegraphics[scale=0.5]{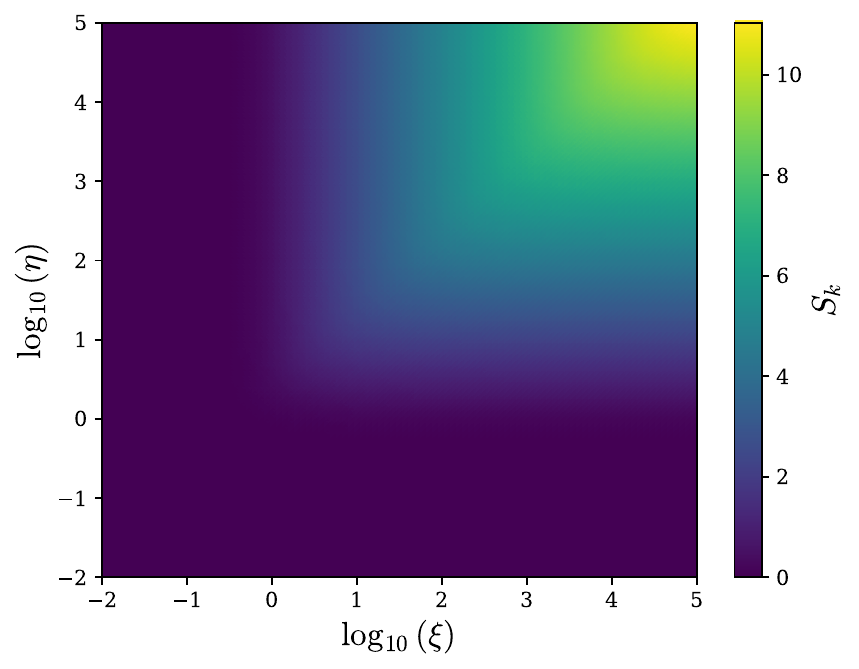}
				\caption{Pairwise entanglement entropy $S_{\vec{k}}$ in the Sauter-Schwinger effect for the IR mode $\omega_{\vec{k}}^+=\mu$.}
		\label{fig:sk}
	\end{center}
\end{figure}

To better highlight the trends in pairwise entanglement, we show them in \ref{fig:sk}. We see that the pairwise entropy for the IR mode freezes in Regimes A~($\eta\gg\xi$) and B~($\eta\ll\xi$) maintaining an ``L-shaped" symmetry consistent with \ref{eq:regimes}, while also growing along the intermediate region $\xi/\eta\sim O(1)$. An alternate measure of pairwise entanglement is available in the form of logarithmic negativity~\cite{2002Vidal.WernerPRA,2005PlenioPRL}, which can serve as a genuine entanglement witness not just for an initial vacuum state but also for Gaussian mixed states. See for instance \cite{2025Kranas.etal} (which appeared during the preparation of this manuscript) where the effects of an initial thermal occupation on pairwise entanglement have been characterized for the Schwinger effect. In this work, we study entropy to capture the salient features of Sauter-Schwinger effect under a unitary evolution, and in particular, the resulting spatial entanglement structure.

\subsection{Real/Imaginary Partition}
From the previous section it is clear that the calculation of entanglement depends on how we partition the two-mode system using different quadrature bases. Let us now consider a third basis, that serves to partition the real and imaginary components of the field, i.e., $\hat{\Phi}_{\vec{k}}=(\hat{\varphi}_1+i\hat{\varphi}_2)/\sqrt{2}$ and $\hat{\Pi}_{\vec{k}}=(\hat{\pi}_1+i\hat{\pi}_2)/\sqrt{2}$. However, to make the covariance matrix elements dimensionless, we will instead choose to rescale the operators with an arbitrary length scale $\epsilon$ across all $\vec{k}$-sectors:
\begin{align}\label{eq:RIbasis}
\sqrt{\mathcal{V}}\vec{Q}&=\{\epsilon^{-1/2}\hat{\varphi}_{1},\epsilon^{-1/2}\hat{\varphi}_{2},\epsilon^{1/2}\hat{\pi}_{1},\epsilon^{1/2}\hat{\pi}_{2}\}
\end{align}
In this basis, the covariance matrix takes the GHDM form similar to \ref{eq:GHDM1}:
\begin{equation}
    \mathbfs{\Sigma}=\begin{bmatrix}
       \tilde{\mathbfs{\Sigma}}_{\varphi\varphi}  &\tilde{\mathbfs{\Sigma}}_{\varphi\pi}\\
       \tilde{\mathbfs{\Sigma}}_{\varphi\pi} &\tilde{\mathbfs{\Sigma}}_{\pi\pi}
    \end{bmatrix},
\end{equation}
where the elements are related to those of \ref{eq:GHDM1} as follows: 
\begin{subequations}\label{eq:Sigma_tilde}
\begin{align}
(\tilde{\Sigma}_{\varphi\varphi})_{\vec{k}\vec{k}'}&=\frac{(\Sigma_{\varphi\varphi})_{\vec{k}\vec{k}'}}{\omega_{\vec{k}}^+\epsilon};\quad\tilde{\mathbfs{\Sigma}}_{\varphi\pi}=\mathbfs{\Sigma}_{\varphi\pi};\\
\quad(\tilde{\Sigma}_{\pi\pi})_{\vec{k}\vec{k}'}&=\omega_{\vec{k}}^+\epsilon(\Sigma_{\pi\pi})_{\vec{k}\vec{k}'}
\end{align}    
\end{subequations}
The real/imaginary (or R/I in short) partitions themselves factorize into pure states, and are each described by the reduced covariance matrix:
\begin{equation}\label{eq:RIcov}
    \mathbfs{\Sigma}_{R,I}=\begin{bmatrix}
       \tilde{\mathbfs{\Sigma}}_{\varphi\varphi} &\tilde{\mathbfs{\Sigma}}_{\varphi\pi}  \\
       \tilde{\mathbfs{\Sigma}}_{\varphi\pi} &\tilde{\mathbfs{\Sigma}}_{\pi\pi} 
    \end{bmatrix}\,.
\end{equation}
The R/I sectors remain unentangled throughout the evolution. This basis can be utilized when a new length scale is introduced into the system, for instance in the form of lattice spacing that helps regulate divergent spatial correlations of the quantum field. While pairwise entanglement for the particles produced in the final state can be calculated in the basis defined in \ref{eq:BDbasis}, we will see in the next section that the basis defined in \ref{eq:RIbasis} becomes natural in the calculation of entanglement between spatial subregions of the field. 

\section{Sauter-Schwinger effect and real-space entanglement}\label{sec:realspace_detailed}

\begin{figure*}[t]
	\begin{center}
		\subfloat[]{%
			\includegraphics[width=0.4\textwidth]{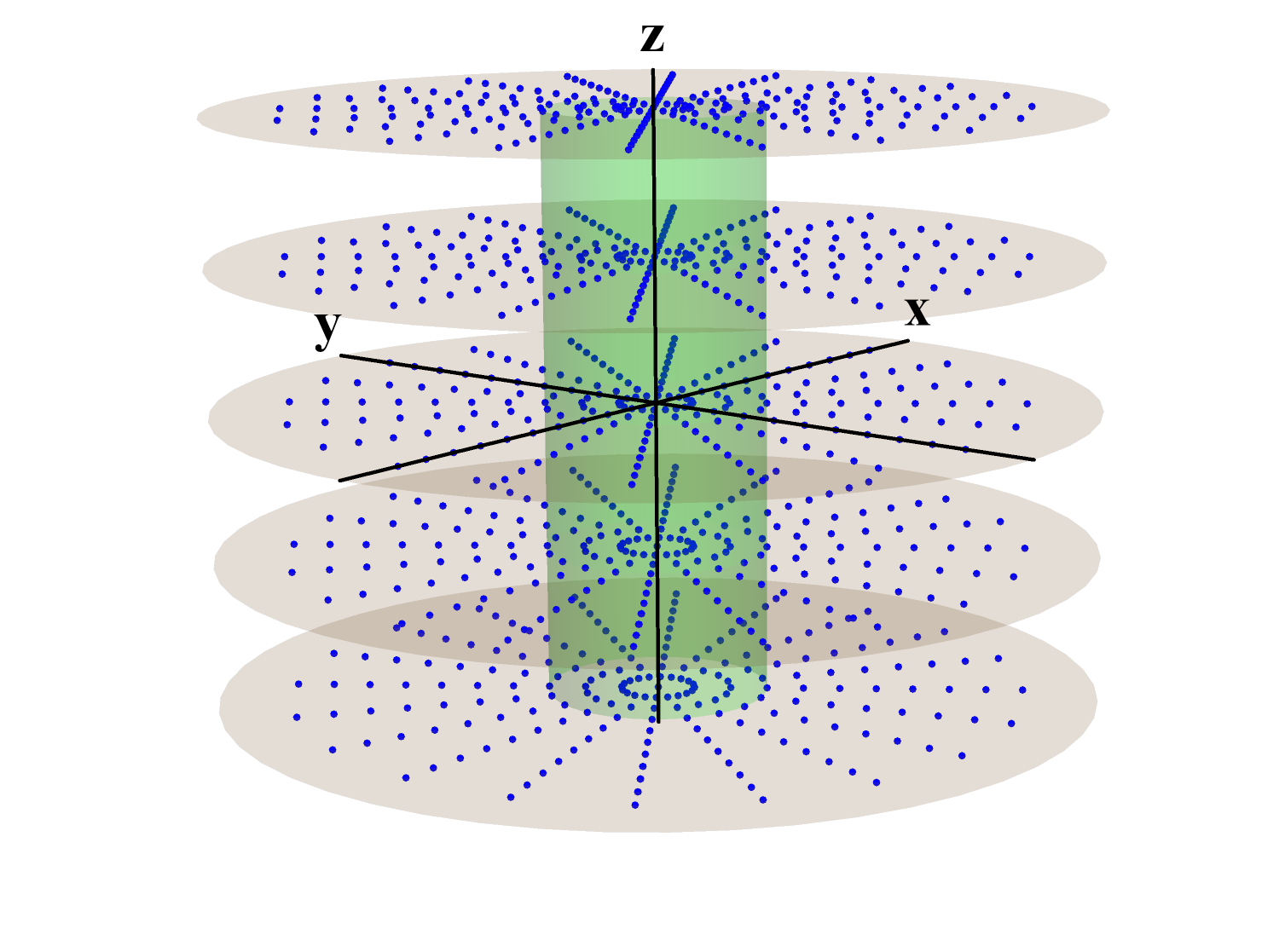}
		}
		\subfloat[]{%
			\includegraphics[width=0.4\textwidth]{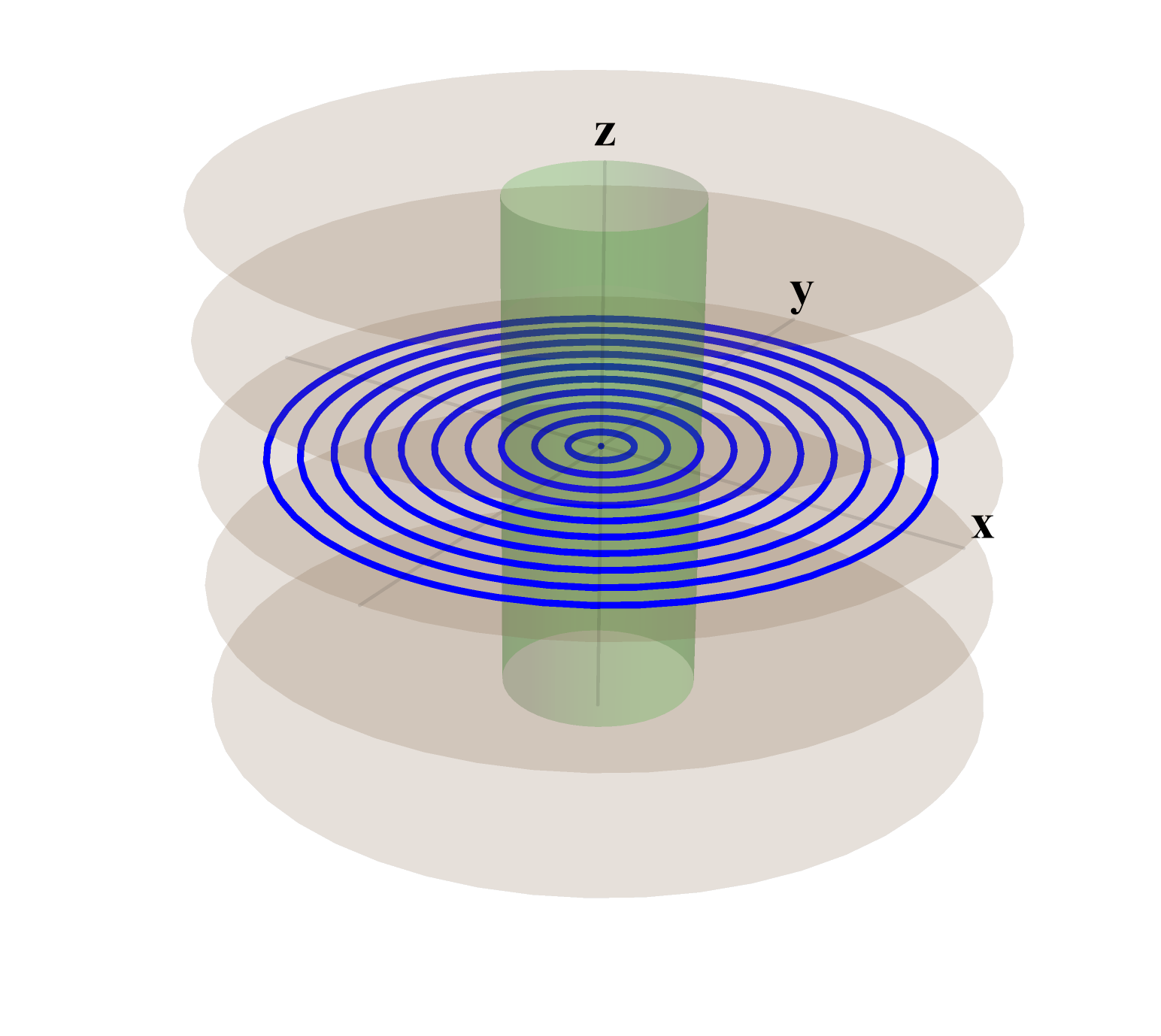}
		}
		\caption{\textit{Left:} Conventional lattice regularization in cylindrical coordinates. \textit{Right:} Dimensionally reduced representation, yielding an effective 1+1-dimensional lattice theory.}
        \label{fig:3plus1_to_1plus1}
	\end{center}
\end{figure*}

Building on the analysis of pair creation in a Sauter pulse and the associated momentum-space entanglement, we now focus on entanglement between spatial subregions. In this section, we present the formalism for quantifying spatial entanglement in scalar QED and provide explicit calculations. We begin with the action for a charged scalar field coupled to an electromagnetic field specified by a vector potential $A^{\mu}$:
\begin{equation}\label{eq:action}
    \mathscr{S}=\int d^4x\left[(D^{*\nu}\Phi^*)(D_{\nu}\Phi)-\mu^2|\Phi|^2\right]
\end{equation}
where $D_{\mu}=\partial_{\mu}+ ieA_{\mu}$ denotes the covariant derivative. For a Sauter-Schwinger pulse along the z-direction, i.e., $A_{\mu}=\delta^{z}_{\mu} A_{z}(t)$, the Lagrangian reduces to:
\begin{multline}\label{eq:lagrangian}
   \mathscr{L}=\int d^3 \vec{x}\Big[|\dot{\Phi}|^2-|\nabla \Phi|^2+ieA_{z}(t)\left(\Phi^*\partial_z\Phi-\Phi\partial_z\Phi^*\right)\\-\left(\mu^2+e^2A_{z}^2(t)\right)|\Phi|^2\Big]
\end{multline}
To perform lattice-regularization and implement the standard techniques for calculating real-space entanglement~\cite{1986Bombelli.etalPRD,1993SrednickiPRL}, we consider a series of decompositions tailored to the problem at hand. 

First, given the cylindrical symmetry of the system, it is convenient to perform a Fourier decomposition along the z-direction as follows:
\begin{equation}
 \Phi(\mathbfs{\varrho},z,t)=\frac{1}{\sqrt{2\pi}}\sum_{l} \,\Phi_{l}(\mathbfs{\varrho},t) e^{\frac{2i\pi l}{L}z},
\end{equation}
which gives us the following Lagrangian where the transverse modes have been decoupled:
\begin{subequations}
\begin{align}
&\mathscr{L}=\sum_{l} \int d^2\mathbfs{\varrho}\,\left[|\dot{\Phi}_{l}|^2-|\nabla_{\mathbfs{\varrho}} \Phi_{l}|^2-\Omega^2|\Phi_{l}|^2\right]\\
&\Omega^2=\mu^2+\left(\frac{2\pi l}{L}+eA_{z}\right)^2\,,
\end{align}   
\end{subequations}
where $\Omega(t)$ takes the form of a time-dependent effective mass.

Second, we decompose the real and imaginary components of the complex scalar field as $\Phi_{l}=(\varphi_{1,l}+i\varphi_{2,l})/\sqrt{2}$, which factorizes the Lagrangian as $\mathscr{L}=\sum_{i=1,2} \mathscr{L}^{(i)}$ where:
\begin{equation}
    \mathscr{L}^{(i)}=\sum_{l}\int  \frac{d^2\mathbfs{\varrho}}{2}\left[\dot{\varphi}_{i,l}^2-(\nabla_{\mathbfs{\varrho}} \varphi_{i,l})^2-\Omega^2\varphi_{i,l}^2\right]
\end{equation}
Each component Lagrangian $\mathscr{L}^{(i)}$ thus describes a real scalar field with an effective time-dependent mass, and all components obey identical equations of motion. Consequently, we can omit the component index $i$ and account for the total entanglement entropy by doubling the contribution from a single component.

Third, we perform an angular decomposition exploiting the axial symmetry of the system:
\begin{subequations}
\begin{align}
    \varphi_{l}(\varrho,\theta,t)&=\sum_{m=-\infty}^{\infty}\frac{\tilde{\varphi}_{lm}(\varrho,t)\cos{m\theta}}{\sqrt{ \pi \varrho(1+\delta_{m,0})}}\\\quad \dot{\varphi}_{l}(\varrho,\theta,t)&=\sum_{m=-\infty}^{\infty}\frac{\dot{\tilde{\varphi}}_{lm}(\varrho,t)\cos{m\theta}}{\sqrt{ \pi \varrho(1+\delta_{m,0})}}
\end{align}
\end{subequations}
Integrating out the $\theta$ component, the Lagrangian reduces to: 
\begin{equation}
    \mathscr{L}=\sum_{lm}\int \frac{d\varrho}{2} \left[\dot{\tilde{\varphi}}_{lm}^2 -\varrho\left\{\partial_\varrho\left(\frac{\tilde{\varphi}_{lm}}{\sqrt{\varrho}}\right)\right\}^2-\Omega^2\tilde{\varphi}_{lm}^2\right]
\end{equation}
This reduction transforms the original 3+1 dimensional problem into a set of independent 1+1 dimensional radial systems labeled by $(l,m)$, with each mode capturing the dynamics along the radial direction, and the angular dependence fully encoded in $m$ (see \ref{fig:3plus1_to_1plus1}). Leveraging this simplification, we shall now construct a lattice formulation that is far 
simpler than a full 3+1 dimensional lattice, enabling faster numerical computation of entanglement entropy for cylindrical spatial bipartitions.

For each $l$ and $m$-mode, the field degrees of freedom can be discretized along the radial direction via
\begin{subequations}
\begin{align}
    \varrho&\rightarrow \varrho_{j}=j\epsilon\quad;\quad 
    1\leq j\leq N_\varrho\,,\\
    \tilde{\varphi}_{lm}(\varrho)&\rightarrow \tilde{\varphi}_{lm,j}\equiv\tilde{\varphi}_{lm}(\varrho_{j})\,,
\end{align}    
\end{subequations}
where $\epsilon$ is the lattice spacing and 
$N_\varrho$ sets the total number of radial lattice points, to obtain the following Hamiltonian:
\begin{align}
H&=\frac{ 1}{2\epsilon}\sum_{lmj}\Bigg[\tilde{\pi}_{lm,j}^2+\left(j+\frac{1}{2}\right)\left\{ \frac{\tilde{\varphi}_{lm,j}}{\sqrt{j}}-\frac{\tilde{\varphi}_{lm,j+1}}{\sqrt{j+1}}\right\}^2\nonumber\\&\qquad\qquad\qquad-\left\{\epsilon^2\Omega^2+\frac{m^2}{j^2}\right\}\tilde{\varphi}_{lm,j}^2\Bigg]\,.
\end{align}
To simplify further, we perform a canonical transformation $\tilde{\pi}_{lm,j}=\epsilon^{1/2}\pi_{lm,j}$ and $\tilde{\varphi}_{lm,j}=\epsilon^{-1/2}\varphi_{lm,j}$, to arrive at a Hamiltonian describing a system of harmonic oscillators with nearest-neighbor coupling:
\begin{multline}
    H=\sum_{lmj}\Bigg[\frac{\pi_{lm,j}^2}{2}+\frac{\left(j+\frac{1}{2}\right)}{2\epsilon^2}\left\{ \frac{\varphi_{lm,j}}{\sqrt{j}}-\frac{\varphi_{lm,j+1}}{\sqrt{j+1}}\right\}^2\\-\frac{1}{2}\left\{\Omega^2+\frac{m^2}{j^2\epsilon^2}\right\}\tilde{\varphi}_{lm,j}^2\Bigg]\,.
\end{multline}
Our aim now is to calculate the entanglement entropy from the above Hamiltonian. 

Being a symplectic invariant measure, the entanglement entropy for spatial bipartitions is independent of the choice of canonical field variables, whether $\{\tilde{\pi}_{lmj},\tilde{\varphi}_{lmj}\}$ or $\{\pi_{lmj},\varphi_{lmj}\}$, i.e.~\cite{2020Grain.VenninJoCaAP,2020Chandran.ShankaranarayananPRD},
\begin{equation}
   H=\epsilon^{-1}\tilde{H} \implies S=\tilde{S}\,.
\end{equation}
It is however convenient to work with the dimensionless Hamiltonian $\tilde{H}=\epsilon H$ for a number of reasons --- first, since the system $\tilde{H}$ is fully characterized by dimensionless parameters which now also include:
\begin{equation}
    \tilde{\mu}=\mu \epsilon;\quad \tilde{t}=\frac{t}{\epsilon};\quad \tilde{\tau}=\frac{\tau}{\epsilon}\,,
\end{equation}
corresponding to the dimensionless mass, time coordinate, and pulse width, respectively. This formulation allows all results to be expressed independently of a specific choice of $\epsilon$. Second, the UV divergence arising from $\epsilon^{-1}\to \infty$ is \textit{folded} into a zero-mode divergence in the rescaled Hamiltonian $\tilde{H}$~\cite{2014Mallayya.etalPRD,2020Chandran.ShankaranarayananPRD}, making the numerical simulations easier to handle and ensuring that dominant contributions to entanglement are always confined to the IR sector. Lastly, the covariance matrix elements are automatically dimensionless, and there is no ambiguity in the choice of quadrature basis $\vec{Q}$. 

We also note that while entanglement entropy is UV-sensitive, the scaling law as such is independent of the details of regularization --- thus making it possible for us to numerically extract the correct leading order behavior from simply a finite number of modes. In fact, for dense enough lattice-points, the framework adopted here can resolve up to subleading terms obtained via the replica trick~\cite{2021Jain.etalPRD}.

\subsection{Boundary conditions and Axial modes}

As mentioned earlier, the Hamiltonian $H$ takes the form of a system of coupled oscillators. In particular, by decomposing the (rescaled) Hamiltonian into each $(lm)$ sector, $\tilde{H}=\sum_{lm}\tilde{H}_{lm}$, we have
\begin{equation}
    \tilde{H}_{lm}=\frac{1}{2}\sum_{i=1}^{N_\varrho}\left[\tilde{\pi}_i^2+\sum_{j=1}^{N_\varrho}K_{ij}(l,m,\tilde{t})\tilde{\varphi}_i\tilde{\varphi}_j\right],
\end{equation}
where $K_{ij}$ is the coupling matrix and the $(lm)$ indices of the field variables have been suppressed for readability. The coupling matrix, in turn, can be separated into time-independent and time-dependent matrices as follows:
\begin{equation}
    \mathbfs{K}(l,m,\tilde{t})=\bar{\mathbfs{K}}(m)+\epsilon^2\Omega^2(l,\tilde{t})\mathbfs{\mathbb{I}}\,,
\end{equation}
where $\mathbb{I}$ is the identity matrix. We see that $\mathbfs{K}$ and $\bar{\mathbfs{K}}$ are simultaneously diagonalizable. 

To obtain a well-defined matrix $\mathbfs{K}$ (or, equivalently, $\bar{\mathbfs{K}}$) the system must be supplemented with appropriate boundary conditions for the discretized field variables. By imposing the Dirichlet boundary conditions at $\varrho=0$ and $\varrho=R=(N_\varrho+1)\epsilon$, we get the following non-zero entries for the coupling matrix $\mathbfs{\bar{K}}$:
\begin{subequations}\label{eq:Kmatrix}
     \begin{align}
		\bar{K}_{jj}&=\frac{m^2}{j^2}+2\,,\\
		\bar{K}_{j,j+1}&=-\frac{j+\frac{1}{2}}{\sqrt{j(j+1)}}=\bar{K}_{j+1,j}\,.
	\end{align}
\end{subequations}
Following the standard procedure for coupled oscillators, the system can be diagonalized in the normal mode basis. To connect with the exact continuum case discussed in Sec.~\ref{subsec:modes_continuum}, we retain the shorthand $\vec{k}=(n,l,m)$ to also denote the normal mode frequencies $\tilde{\omega}_\vec{k}$ of the discretized problem. A similar analysis then leads to
\begin{align} \tilde{\omega}_\vec{k}^2(\tilde{t})&=\tilde{\kappa}_{nm}^2+\tilde{p}_l^2(\tilde{t})+\tilde{\mu}^2\,,
\end{align}
where the adiabatic transverse momenta are defined as
\begin{equation}
    \tilde{p}_l(\tilde{t})=\frac{2\pi l\epsilon}{L}-\tilde{\mu}\xi\left(1+\tanh{\frac{\tilde{t}}{\tilde{\tau}}}\right)\,.
\end{equation}
The former equation is nothing but the direct analogue of \ref{eq:normal_mode_continuum}. Here, $\tilde{\kappa}_{nm}$ is the $n$-th eigenvalue of $\mathbfs{\bar{K}}(m)$, and all the parameters with a tilde on top are dimensionless upon rescaling with $\epsilon$.

So far, our discussion of the discretization has closely paralleled the continuum analysis. We now turn to one of the first subtleties that arises specifically in the numerical implementation of the discretized problem. For the transverse tower of modes, recall that we have a discrete spectrum from fixing periodic boundary conditions at $z=\pm L/2$, indexed by $l\in \mathbb{Z}$. In practice, numerical computations necessitate a careful regularization of the formally infinite sum over these modes to a finite subset $\tilde{\mathbb{Z}}\subset \mathbb{Z}$. Moreover, due to the explicit time dependence of the system, this regularization may itself be time dependent in principle, i.e $\tilde{\mathbb{Z}}\rightarrow \tilde{\mathbb{Z}}_{\tilde{t}}$.

Our regularization strategy is guided by the following considerations. First, at each time slice, the infrared modes of the rescaled Hamiltonian $\tilde{H}$ --- which dominate the entanglement entropy --- must always be retained in order to correctly extract the leading-order scaling behavior. A straightforward way to ensure this is to recenter the median value of the \textit{regularized} set of adiabatic momenta $\{\tilde{p}_l(\tilde{t})|l\in\tilde{\mathbb{Z}}_{\tilde{t}}\}$ to zero at each time slice, while keeping the cardinality of $|\tilde{\mathbb{Z}}_{\tilde{t}}|\equiv N_{z}+1$ fixed for consistency across time --- the same cardinality as that of the spatial transverse modes fixed via the boundary condition $L=N_z\epsilon$. 
The momenta are therefore spaced apart by the IR cutoff as $\Delta p_l=2\pi/L$, and bandlimited by the UV cutoff as $|p_l|\leq \pi/\epsilon$. For the mode functions at a particular time-slice, we hence plug in the following set of adiabatic momenta $\tilde{p}_l(\tilde{t})$ evolving from their respective initial values
$\tilde{p}_l^-$ as:
\begin{subequations}
\begin{align}
    \tilde{p}_l(\tilde{t})&=\tilde{p}_l^--\tilde{\mu}\xi\left[1+\tanh{\frac{\tilde{t}}{\tilde{\tau}}}\right]\\
    &=\frac{2\pi l}{N_z}; \quad l=0,\pm 1,\pm 2,.. , \pm \frac{N_z}{2}\,.
\end{align}    
\end{subequations}
This ensures that at each time-slice the dominant contributions to entanglement entropy are counted, allowing us to numerically extract the correct leading order scaling law from a finite number of modes, otherwise only accessible in the continuum limit. Particularly for the final state, the above choice corresponds to $\tilde{p}_l^+=\tilde{p}_l^--2\tilde{\mu}\xi=2\pi l/N_z$.

Lastly, the parameters are chosen such that the axial and radial extents of the system remain much larger than the Compton wavelength of the particle, i.e., $L,R\gg \mu^{-1}$. This ensures adequate resolution of the long-range, nonlocal correlations emerging from pair creation. In terms of the dimensionless parameters, this condition translates to:
\begin{equation}\label{eq:resolution}
    \tilde{\mu}^{-1}\ll N_\varrho,\,N_z\,.
\end{equation}
Accordingly, for the simulations in this paper, we choose $\tilde{\mu}=1$, $N_\varrho=300$ and $N_z=100$ unless otherwise specified. Although one could adopt a finer lattice ($\tilde{\mu}\ll1$), doing so would require a substantial increase in $N_z$ and $N_\varrho$ to maintain \ref{eq:resolution}, significantly increasing the computational burden without affecting the dominant entanglement scaling behavior.

\begin{figure*}[t]
	\begin{center}
		\subfloat[][$\xi=10^{-2},\,\eta=10$]{%
			\includegraphics[width=0.25\textwidth]{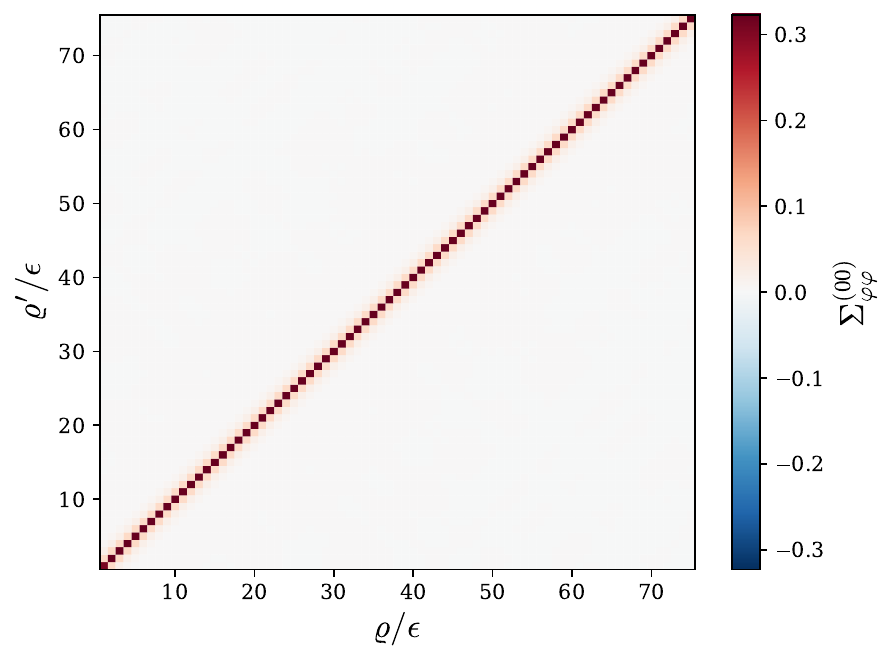}\label{CP1}
			\includegraphics[width=0.25\textwidth]{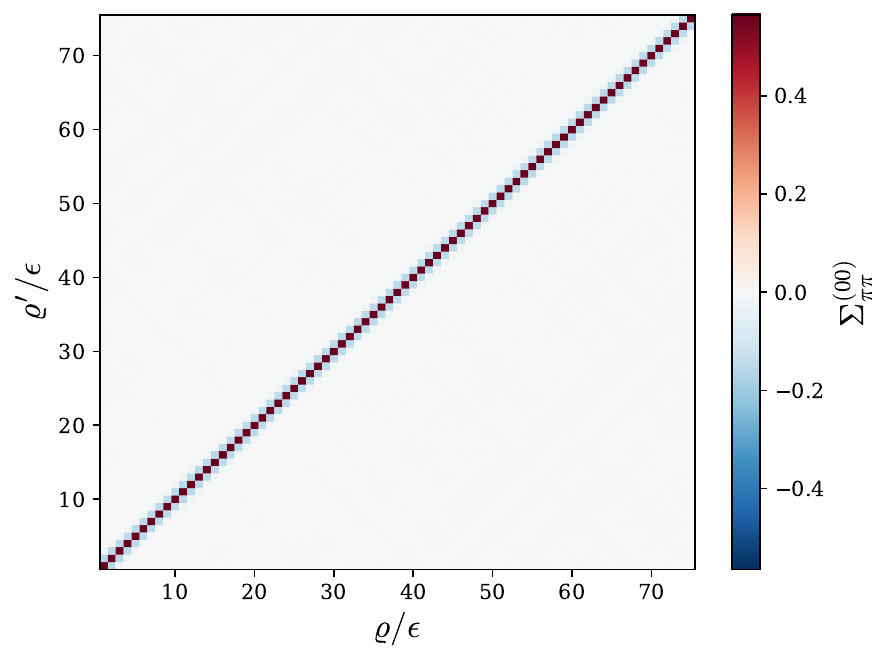}\label{CP2}
		}
		\subfloat[][$\xi=10^{2},\,\eta=10$]{%
			\includegraphics[width=0.25\textwidth]{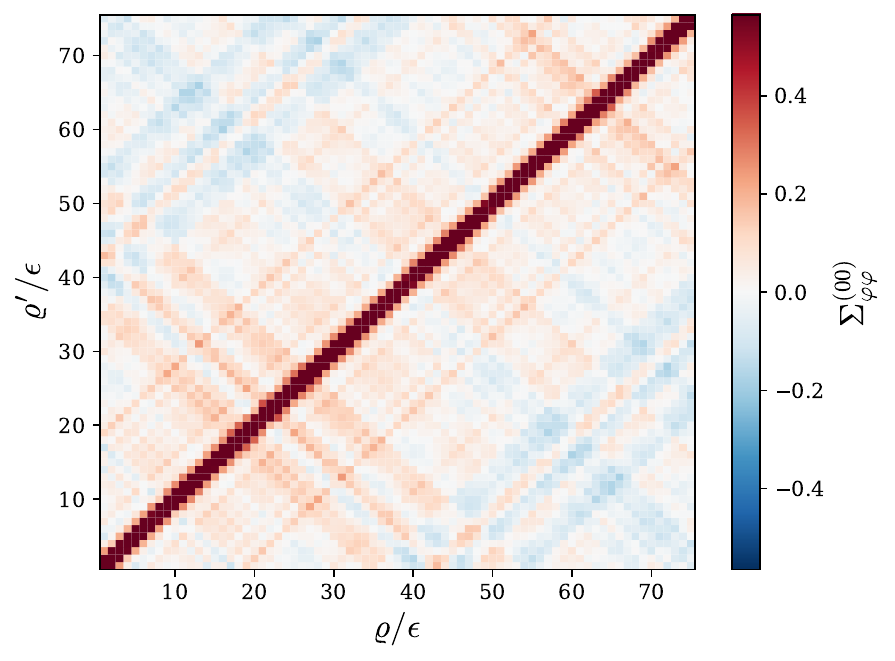}\label{CP3}
			\includegraphics[width=0.25\textwidth]{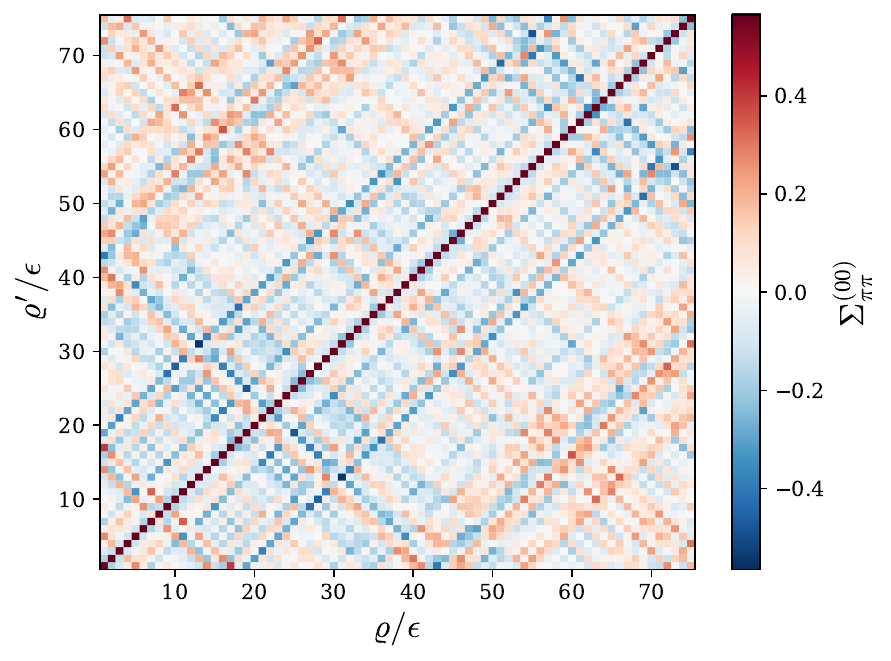}\label{CP4}
		}
		\caption{The final state ($t/\tau =10^{20}$) correlation structure for the $l=m=0$ sector in (a) the weak-field regime, and (b) the strong-field regime. The emergence of nonlocal correlation peaks in the latter indicates pair creation due to Sauter-Schwinger effect. Here, $\tilde{\mu}=1$ and $N_\varrho=300$.}
		\label{fig:corrpeaks}
	\end{center}
\end{figure*}

\subsection{Covariance matrix and nonlocal correlations}

The quadratic form of the Hamiltonian $H_{lm}$ implies that each normal mode evolves unitarily from a harmonic oscillator ground state defined at $\tilde{t}\to-\infty$, preserving its Gaussian form~\cite{2008LoheJoPAMaT}:
\begin{multline}\label{eq:Psi_k}
    \Psi_\vec{k}\left[\tilde{\varphi}_{\vec{k}},\tilde{t}\right]=\left(\frac{\tilde{\omega}_\vec{k}^-}{\pi \Upsilon_\vec{k}^2}\right)^{1/4}\exp\Bigg[-\left(\frac{\tilde{\omega}_\vec{k}^-}{\Upsilon_\vec{k}^2}-\frac{i\partial_{\tilde{t}}\Upsilon_\vec{k}}{\Upsilon_\vec{k}}\right)\frac{\tilde{\varphi}_\vec{k}^2}{2}\\-\frac{i\omega_\vec{k}}{2}\int \frac{dt}{b_{\vec{k}}^3}\Bigg],
\end{multline}
where $\tilde{\omega}_{\vec{k}}^-=\epsilon\omega_{\vec{k}}^-$ is the initial mode frequency, and the scaling parameter $\Upsilon_\vec{k}$ is the solution to the nonlinear Ermakov-Pinney equation~\cite{1950PinneyPotAMS}:
\begin{equation} \partial_{\tilde{t}}^2\Upsilon_\vec{k}(\tilde{t})+\tilde{\omega}_\vec{k}^2(\tilde{t})\Upsilon_\vec{k}(\tilde{t})=\frac{\left(\tilde{\omega}_\vec{k}^-\right)^2}{\Upsilon_\vec{k}^3(\tilde{t})}
\end{equation}
The scaling parameters are related to the mode-functions defined in \ref{eq:modefn} as follows:
\begin{equation}\label{eq:ermakovconnection}
    \Upsilon_\vec{k}=\sqrt{2\omega_\vec{k}^-}|v_\vec{k}|;\,\,\, \partial_{\tilde{t}}\Upsilon_\vec{k}=\frac{\sqrt{2\omega_{\vec{k}}^-}\Re\left[v_{\vec{k}}\partial_{\tilde{t}}{v}_{\vec{k}}^*\right]}{|v_{\vec{k}}|}
\end{equation}
This way, \ref{eq:Psi_k} provides a complete description of the quantum dynamics of the vacuum. We now describe how to extract information about the correlations encoded in this state and, ultimately, compute the geometric entanglement entropy.

The Gaussian form of the wavefunction allows us to read off the \textit{time-dependent} covariance matrix for each $\vec{k}$-mode subsystem:
\begin{align}
    \mathbfs{\Sigma}_{\vec{k}}&=\frac{1}{2}\begin{bmatrix}
        \langle\{\hat{\tilde{\varphi}}_\vec{k},\hat{\tilde{\varphi}}_\vec{k}\}\rangle &\langle\{\hat{\tilde{\varphi}}_\vec{k},\hat{\tilde{\pi}}_\vec{k}\}\rangle  \\
       \langle\{\hat{\tilde{\varphi}}_\vec{k},\hat{\tilde{\pi}}_\vec{k}\}\rangle& \langle\{\hat{\tilde{\pi}}_\vec{k},\hat{\tilde{\pi}}_\vec{k}\}\rangle 
    \end{bmatrix}=\begin{bmatrix}
        \tilde{\mathbfs{\Sigma}}_{\varphi\varphi} &\tilde{\mathbfs{\Sigma}}_{\varphi\pi}\\
      \tilde{\mathbfs{\Sigma}}_{\varphi\pi}&\tilde{\mathbfs{\Sigma}}_{\pi\pi}
    \end{bmatrix}\,,
\end{align}
where the block-elements are explicitly given by
\begin{align}
        (\tilde{\Sigma}_{\varphi\varphi})_{\vec{k}\vec{k}'} &=\frac{\Upsilon_\vec{k}^2}{2\tilde{\omega}_{\vec{k}}^-}\delta_{\vec{k}\vec{k}'};\,\,\, (\tilde{\Sigma}_{\varphi\pi})_{\vec{k}\vec{k}'} =\frac{\Upsilon_\vec{k}\partial_{\tilde{t}}\Upsilon_\vec{k}}{2\tilde{\omega}_{\vec{k}}^-}\delta_{\vec{k}\vec{k}'};\nonumber\\
        (\tilde{\Sigma}_{\pi\pi})_{\vec{k}\vec{k}'} &=\left[\frac{\tilde{\omega}_{\vec{k}}^-}{2\Upsilon_\vec{k}^2}+\frac{\left(\partial_{\tilde{t}}\Upsilon_\vec{k}\right)^2}{2\tilde{\omega}_{\vec{k}}^-}\right]\delta_{\vec{k}\vec{k}'}
\end{align}
It can be readily verified, using \ref{eq:ermakovconnection}, that 
the single-mode covariance matrix exactly coincides with that of the R/I subsystem defined in \ref{eq:RIcov}, where the length scale introduced there now corresponds to the UV regulator $\epsilon$. 
In the asymptotic future $\tilde{t}\to\infty$, the matrix elements further coincide with \ref{eq:Sigma_tilde}, and can be expressed in terms of the squeezing parameters defined in \ref{eq:squeeze_para} as follows:
\begin{align}  
(\tilde{\Sigma}_{\varphi\varphi})_{\vec{k}\vec{k}'}&=\frac{\delta_{\vec{k}\vec{k}'}}{2\tilde{\omega}_{\vec{k}}^+}\left[\cosh{2r_{\vec{k}}}+\sinh{2r_{\vec{k}}}\cos{2\vartheta_{\vec{k}}}\right]\nonumber\\
 (\tilde{\Sigma}_{\varphi\pi})_{\vec{k}\vec{k}'}&=\frac{\delta_{\vec{k}\vec{k}'}}{2}\sinh{2r_{\vec{k}}}\sin{2\vartheta_{\vec{k}}}\\
    (\tilde{\Sigma}_{\pi\pi})_{\vec{k}\vec{k}'}&=\frac{\tilde{\omega}_{\vec{k}}^+\delta_{\vec{k}\vec{k}'}}{2}\left[\cosh{2r_{\vec{k}}}-\sinh{2r_{\vec{k}}}\cos{2\vartheta_{\vec{k}}}\right]\nonumber
\end{align}
This is unsurprising: it simply reflects that a canonical transformation of the standard quadrature variables yields the normal-mode oscillators of the system. The more challenging task, however, is to infer the nontrivial spatial correlations of the vacuum from the seemingly simple block-diagonal covariance matrix in the normal-mode basis, and to follow how these correlations evolve over time. We address this next.

The recipe for this translation involves the matrix $\mathbf{V}_{lm}$ that transforms the discretized field variable from the real-space basis to the normal mode basis:
\begin{align}
    \sum_{n}\tilde{\varphi}_{\mathbfs{k}}(\mathbf{V}_{lm})_{nj}=\tilde{\varphi}_{lm,j}\,\,;\,\, \vec{k}=(l,m,n)\,.
\end{align}
The matrix $\mathbfs{V}_{lm}$ has the defining property that it diagonalizes the coupling matrix as $\mathbfs{V}_{lm}\mathbfs{K}(l,m,\tilde{t}) \mathbfs{V}^{T}_{lm}=\diag\{\tilde{\omega}_\vec{k}^2(\tilde{t})\}$. Given this transformation matrix, the real-space covariance matrix $\mathbfs{\Sigma}^{(lm)}$ can then be constructed for each $lm$-sector as:
\begin{subequations}\label{eq:Sigma_lm}
\begin{align}
    \mathbfs{\Sigma}^{(lm)}&=\begin{bmatrix}
        \mathbfs{V}_{lm}^T\tilde{\mathbfs{\Sigma}}_{\varphi\varphi}\mathbfs{V}_{lm} & \mathbfs{V}_{lm}^T\tilde{\mathbfs{\Sigma}}_{\varphi\pi}\mathbfs{V}_{lm}\\ 
       \mathbfs{V}_{lm}^T\tilde{\mathbfs{\Sigma}}_{\varphi\pi}\mathbfs{V}_{lm} & \mathbfs{V}_{lm}^T\tilde{\mathbfs{\Sigma}}_{\pi\pi}\mathbfs{V}_{lm}
    \end{bmatrix}\,,\\&=\begin{bmatrix}
        \mathbfs{\Sigma}_{\varphi\varphi}^{(lm)} & \mathbfs{\Sigma}_{\varphi\pi}^{(lm)}\\ 
       \mathbfs{\Sigma}_{\varphi\pi}^{(lm)} & \mathbfs{\Sigma}_{\pi\pi}^{(lm)}
    \end{bmatrix}\,.
\end{align}
\end{subequations}
In the exact continuum limit, the analogue of $\mathbfs{V}_{lm}$ is a kernel that may be viewed as a generalization of the Hankel transform to a finite disk $0<\varrho<R$. In the discretized setting, $\mathbfs{V}_{lm}$ can be constructed numerically and subsequently used to evaluate the covariance matrix above.

To illustrate the spatial correlations, we plot the asymptotic covariance matrix $\mathbfs{\Sigma}^{(lm)}$ in the $l=m=0$ sector for two representative choices of the parameters. As shown in \ref{fig:corrpeaks}, the correlation structure of the final state depends sensitively on $(\xi,\eta)$, with nonlocal correlation peaks emerging in the strong-field regimes atop the short-distance vacuum correlations near $\varrho\to \varrho'$ that otherwise characterize $\mathbfs{\Sigma}_{\varphi\varphi}^{(lm)}$ and $\mathbfs{\Sigma}_{\pi\pi}^{(lm)}$.

Such nonlocal corrections are indicative of pair creation, particularly serving as smoking-gun signatures for Hawking radiation~\cite{2008Balbinot.etalPRA,2013Anderson.etalPRD} --- with their recent observation in analogue black hole setups having resulted in the experimental verification of the same~\cite{2016SteinhauerNP,2019MunozdeNova.etalN}. We note that these signatures however differ for the Sauter-Schwinger effect in the following ways --- (i) several nonlocal peaks manifest throughout the system for various values of $\varrho$ in the Sauter-Schwinger effect, whereas for Hawking-radiation a dominant peak emerges from near the horizon ($r\sim r_h$) along the paths of the Hawking quanta, and (ii) for the former the peaks are dynamically shifting corresponding to a free-evolving squeezed state as $\tilde{t}\to\infty$, whereas for the latter the dominant peak is static.

With this preparation, we now proceed to the central objective of 
evaluating the entanglement entropy under a cylindrical bipartition of the system.

\subsection{Scaling and dynamics of entanglement entropy}

In this section, we finally put to use the concepts introduced in \ref{sec:warmup} to compute the geometrical entanglement entropy. Since we have had to describe a variety of covariance matrices, all denoted with some decoration on the symbol $\mathbfs{\Sigma}$, we are going to start with the precise choice, out of these, that is relevant to the discussion here in light of \ref{sec:warmup}. For the proceeding discussion, in order to apply the formalism outlined in \ref{sec:warmup}, we identify
\begin{align}
    \mathbfs{\Sigma}\rightarrow \mathbfs{\Sigma}^{(lm)}
\end{align}
where $\mathbfs{\Sigma}^{(lm)}$ is defined in \ref{eq:Sigma_lm}. Having fixed this choice, we next specify concrete definitions of the “in” and “out” sets of modes.

The central idea is to bipartition the spatial region of the system into interior and exterior domains, following the construction outlined in \ref{sec:warmup}, and to compute the entanglement entropy of the reduced density matrix obtained by tracing out the exterior degrees of freedom. In the continuum limit, the interior (IN-subsystem) region is taken to be
\begin{align}
    \mathcal{D}_{\rm IN}=\{(\varrho,z,\phi)\quad|\quad 0\leq\varrho\leq r\}\quad;\quad r<R\,.
\end{align}
Within the discretized framework introduced in the previous subsection, this region is modeled as
\begin{align}
    \mathcal{D}_{\rm in}=\{j\quad|\quad 
    1\leq j\leq n_{\varrho}\}\quad;\quad n_{\varrho}\leq N_{\varrho}\,,
\end{align}
where we identify the cylinder radius as $r=n_{\varrho}\epsilon$. The exterior (OUT-subsystem) region, analogously, is given by
\begin{align}
    \mathcal{D}_{\rm out}=\{j\quad|\quad N_{\varrho}\geq j> n_{\varrho}\}\quad;\quad n_{\varrho}\leq N_{\varrho}\,.
\end{align}

Recall that at each point in $\mathcal{D}_{\rm in}$ (and similarly in $\mathcal{D}_{\rm out}$) there exists a tower of $(l,m)$ modes. Strictly speaking, the interior region should therefore be understood as the union of the corresponding IN subsystems across all such modes. However, since the $(l,m)$ sectors factorize in the present setup, this distinction is straightforward to handle. In practice, its only effect is that the entanglement entropy — the quantity of ultimate interest — is obtained by summing the contributions from each $(l,m)$ sector. For this reason, we will not belabor this distinction in what follows and will refer to the IN and OUT subsystems without further qualification.

In direct analogy with the construction of \ref{sec:warmup}, we define the reduced covariance matrix associated with the IN subsystem as
\begin{align}
\mathbfs{\sigma}\rightarrow \mathbfs{\sigma}^{(lm)} &=\begin{bmatrix}       \mathbfs{\sigma}_{\varphi\varphi}^{(lm)}&\mathbfs{\sigma}_{\varphi\pi}^{(lm)}\\\mathbfs{\sigma}_{\varphi\pi}^{(lm)T}&\mathbfs{\sigma}_{\pi\pi}^{(lm)}    \end{bmatrix}\,,\\
\left(\sigma_{\varphi\varphi}^{(lm)},\sigma_{\pi\pi}^{(lm)},\sigma_{\varphi\pi}^{(lm)}\right)_{ij}&\equiv\left(\Sigma_{\varphi\varphi}^{(lm)},\Sigma_{\pi\pi}^{(lm)},\Sigma_{\varphi\pi}^{(lm)}\right)_{ij\in\mathcal{D}_{in}}\nonumber
\end{align}
With this identification in place, we 
can now compute the entanglement entropy by directly applying the general prescription outlined in \ref{sec:warmup}, beginning with \ref{eq:def_sigma_in} and culminating in the definition of entanglement entropy given in \ref{eq:Entdef}. 

\begin{figure*}[!htb]
	\begin{center}
		\subfloat[][]{%
			\includegraphics[width=0.46\textwidth]{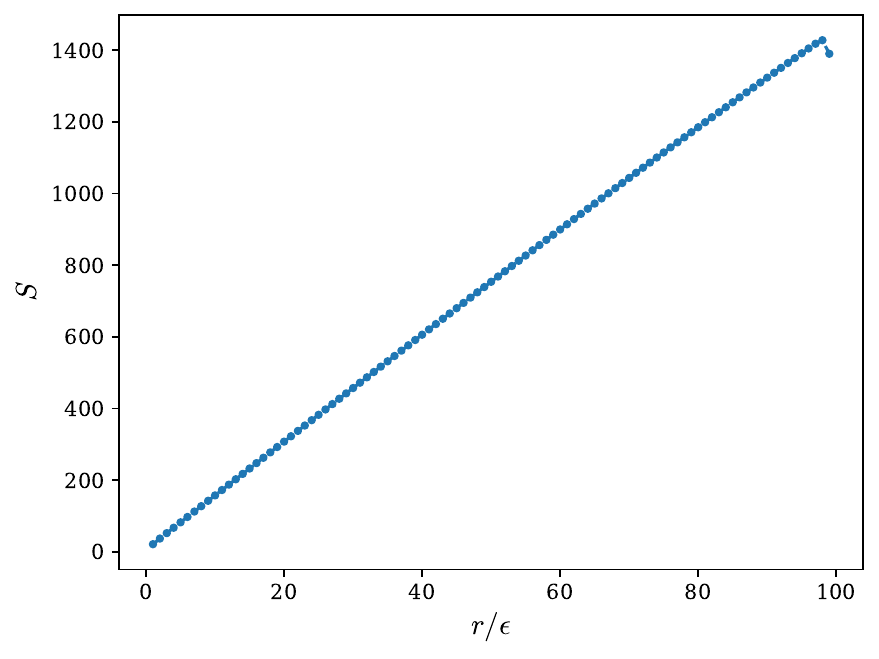}\label{fig:AreaLaw}
		}
		\subfloat[][]{%
			\includegraphics[width=0.45\textwidth]{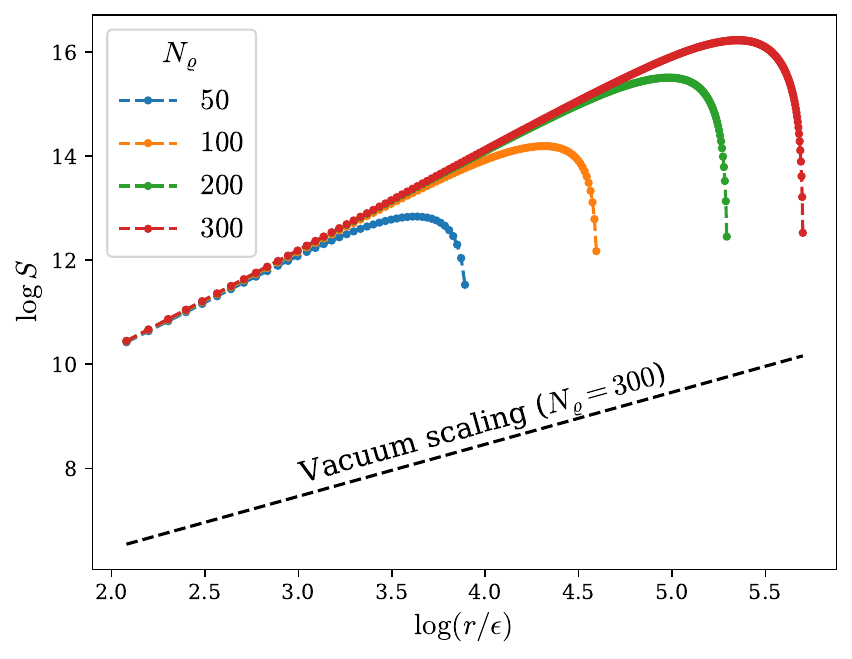}\label{fig:IR_Radial}
		}
		
		\caption{Subsystem scaling of entanglement entropy in (a) the vacuum state for $N_\varrho=100$, and (b) the final state ($t/\tau=10^{20}$) when $\xi=10^2$, $\eta=10$ for various values of $N_\varrho$. Here, we set $\tilde{\mu}=1$ and $N_z=100$.}
		\label{fig:Scaling}
	\end{center}
\end{figure*}

\begin{figure*}[t]
	\begin{center}
		\subfloat[]{%
			\includegraphics[width=0.33\textwidth]{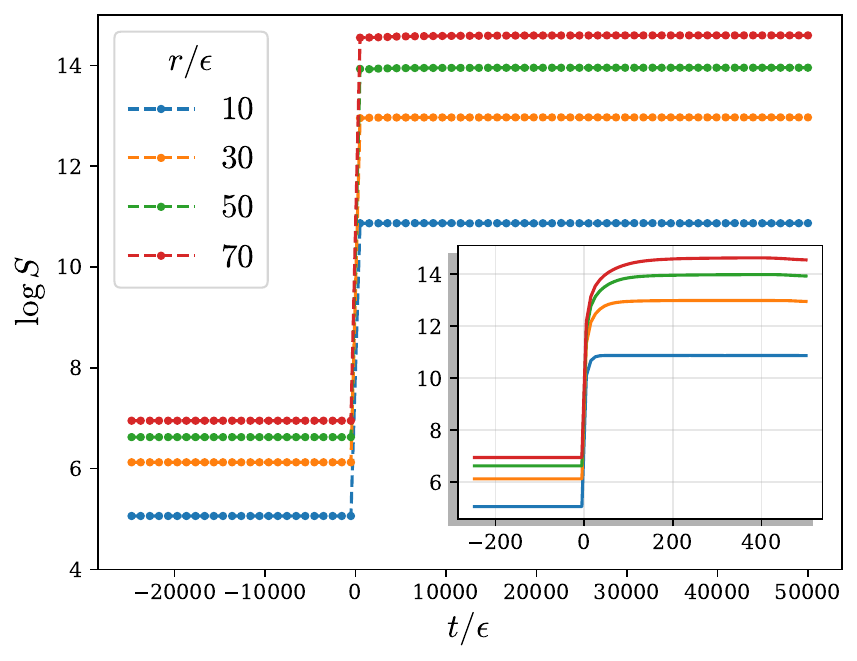}\label{TE1}
		}
		\subfloat[]{%
			\includegraphics[width=0.33\textwidth]{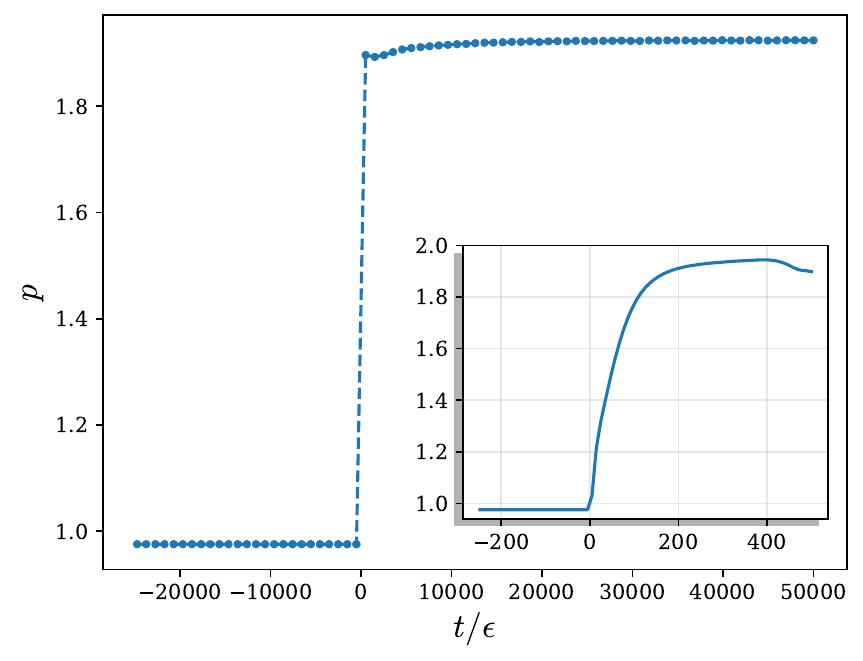}\label{TE2}
		}
		\subfloat[]{%
			\includegraphics[width=0.33\textwidth]{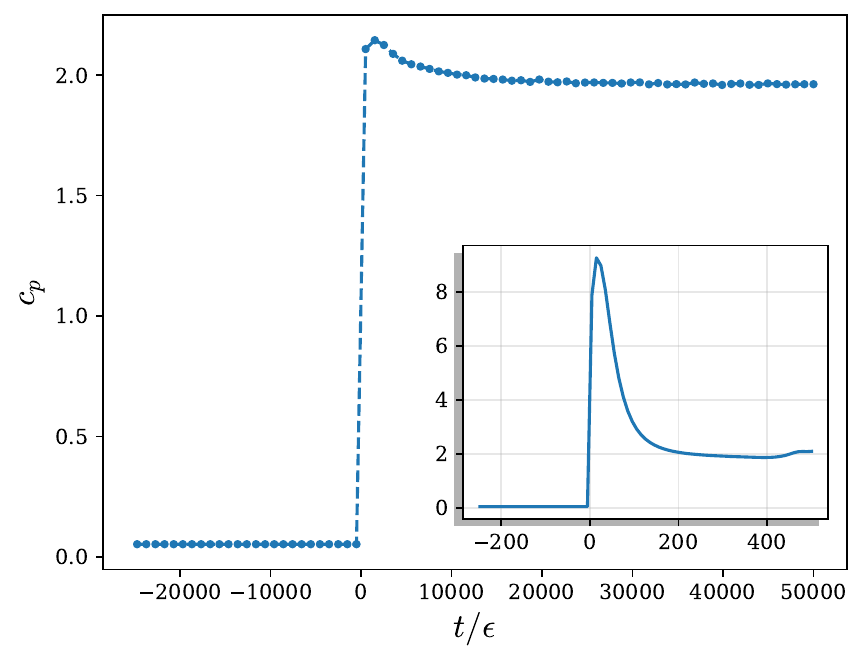}\label{TE3}
		}

		\caption{Time evolution plots of (a) the entanglement entropy for various subsystem sizes, (b) the power-law fit for entropy scaling, and (c) the pre-factor for entropy scaling. Here, we consider $\xi=10^2$, $\eta=10$, $\tilde{\mu}=1$, $N_z=100$, $N_\varrho=300$ and subsystem sizes $R/20<r< R/4$ for the power-law fit.}
		\label{fig:TE}
	\end{center}
\end{figure*}

The outcome of this procedure is a quantity denoted by $S_{lm}(r)$, which represents the contribution of the $(l,m)$ sector to the entanglement entropy of a cylindrical subregion of radius $r = \epsilon\, n_{\varrho}$. The total entanglement entropy 
associated with the same subregion is then given by:
\begin{equation}
    S(r)=2\sum_{lm}S_{lm}(r),
\end{equation}
where the factor of $2$ arises from counting contributions of both the real and imaginary components of the scalar field. Note that for an angular decomposition along a 2D plane, the entanglement entropy converges quite rapidly as $S_{lm}(r)\propto (r/m\epsilon )^4$~\cite{2013Braunstein.etalJoHEP} where $m\gg N_\varrho$ and $\varrho\gg\epsilon$. The summation will therefore be truncated at a maximum value of about $m_{\rm max}\sim 550$ for the numerical simulations employed in this paper, to keep the computational runtimes in check while ensuring convergence to a reasonable degree. For the initial vacuum state at $\tilde{t}\to-\infty$, we observe that the entanglement entropy has a leading order area-law scaling (\ref{fig:Scaling}\subref{fig:AreaLaw}):
\begin{equation}
    S^{-}\sim c\frac{A}{\epsilon^2};\quad A=2\pi r L,
\end{equation}
with a numerically estimated pre-factor of $c\sim 0.046$, which is twice that of a real scalar field~\cite{1986Bombelli.etalPRD,1993SrednickiPRL}. 

\begin{figure*}[!htb]
	\begin{center}
		\subfloat[][]{%
			\includegraphics[width=0.45\textwidth]{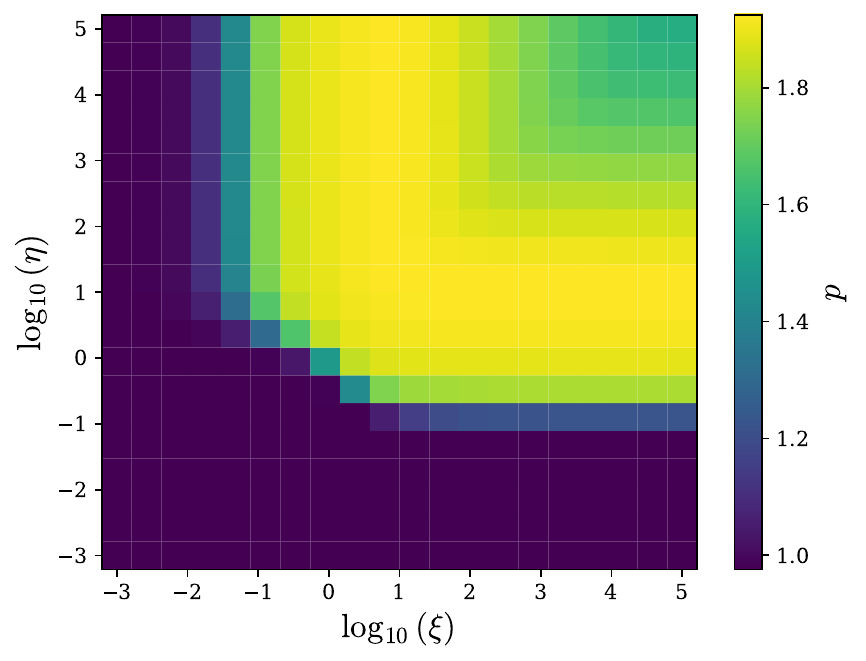}\label{fig:sqa}
		}
		\subfloat[][]{%
			\includegraphics[width=0.452\textwidth]{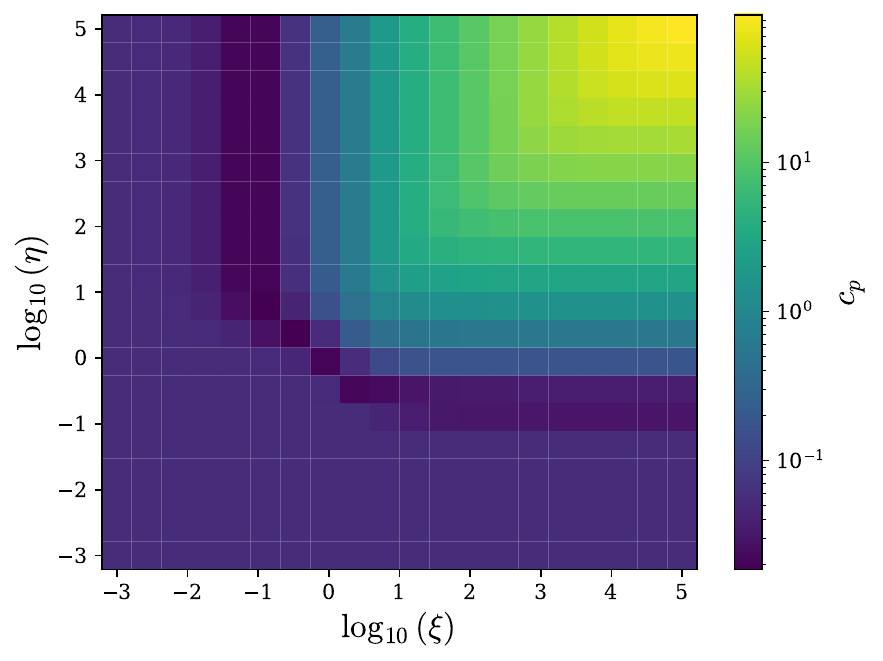}\label{fig:sqb}
		}
		
		\caption{(a) Power law $p$ and (b) pre-factor $c_p$ for the final state entanglement entropy scaling in the Sauter-Schwinger effect. Here $\tilde{\mu}=1$, $N_z=100$, and $N_\varrho=300$. The power-law fits are obtained for $R/20< r< R/4$, with a maximum error of $\Delta p\sim 0.008$, and a minimum \textit{goodness of fit} of $\rm{R}^2\sim 0.999$.}
		\label{fig:SquarePlot}
	\end{center}
\end{figure*}

Upon passing an electric field pulse, the system unitarily evolves to an asymptotic final state where the entropy scaling can potentially deviate from an area-law. Such deviations are expected for certain choices of the parameters $(\xi,\eta)$ where nonlocal correlations (\ref{fig:corrpeaks}) can emerge in the final state via the Sauter-Schwinger effect. One such choice is explored in \ref{fig:Scaling}\subref{fig:IR_Radial}, where we see that the entanglement entropy scales faster than the vacuum case, then dips on approaching the boundary. For typical states, such a turnover is expected when the subsystem volumes are comparable, i.e., at about $r\sim R/\sqrt{2}$ in our setup. We therefore 
confine to subsystem sizes up to a fraction, i.e. $r<R/4$, that is well before the turnover point, and wherein the log-plot is linear. One can then ascribe a \textit{power-law} fitting function for the leading order term in entanglement entropy:
\begin{equation}
    S(\tilde{t})\sim 2\pi N_zc_p(\tilde{t})\left(\frac{r}{\epsilon}\right)^{p(\tilde{t})},
\end{equation}
and track the dynamics of the power-law $p$ and pre-factor $c_p$ --- extracted respectively from the slope and intercept of the $\log{S}$ vs $\log{(r/\epsilon)}$ plot. The power law hence obtained is however slightly underestimated due to (i) the subleading logarithmic corrections to entanglement entropy, and (ii) truncation of the entropy sum which fully converges only as $m\to\infty$. As a result, the area-law corresponding to the vacuum state in \ref{fig:Scaling}\subref{fig:AreaLaw} leads to a fit of only up to $p\sim 0.97$ (a 3\% error) even after excluding the boundary points, and the same can be expected for the power-law fits at later times. For instance, for the final state in \ref{fig:Scaling}\subref{fig:IR_Radial} we obtain $p\sim1.93$, which lies within a $4\%$ error bar from a volume-law scaling. 

The dynamics of entanglement entropy in the Sauter-Schwinger effect leads to some interesting features as observed in \ref{fig:TE}. First, the Sauter-Schwinger pulse amplifies the overall entanglement entropy content between spatial subregions, as well as the power-law/pre-factor for the entropy scaling. The numerical fit is carried out for subsystem sizes in the range $R/20<r< R/4$, where the lower bound truncates spurious finite-size effects near the boundary at $\varrho=0$. Although the correlations become highly oscillatory as the normal mode asymptotes to a freely evolving squeezed state, the final state entanglement entropy is found to be robust to these oscillations --- subsequently freezing the scaling power-law/pre-factor as $t\to\infty$.

Upon performing a sweep of the final state ($t/\tau=10^{20}$) entropy scaling across the ($\xi,\eta$)-parameter space, we make some important observations via \ref{fig:SquarePlot}. First, both the power-law and pre-factor reflect certain salient features of the particle spectrum in the IR sector \ref{eq:regimes} --- mainly, that they become $\eta$-independent in Regime A ($\xi\ll\eta$) and $\xi$-independent in Regime B ($\xi\gg\eta$), revealing an ``L-shaped" symmetry. Second, we observe that the pre-factor grows similarly to the pairwise entanglement content in the IR sector as observed in \ref{fig:sk}, whereas the power-law plot captures finer details of the particle distribution in the IR sector. That is, while the weak-field/adiabatic regime is dominated by area-law states, the scaling-law grows in the strong-field/nonadiabatic regime where the low-energy pairs follow a near-thermal spectrum~\ref{eq:thermality}.

\begin{figure}[!b]
\vspace*{1em}
	\begin{center}
		\centering
	\includegraphics[scale=0.465]{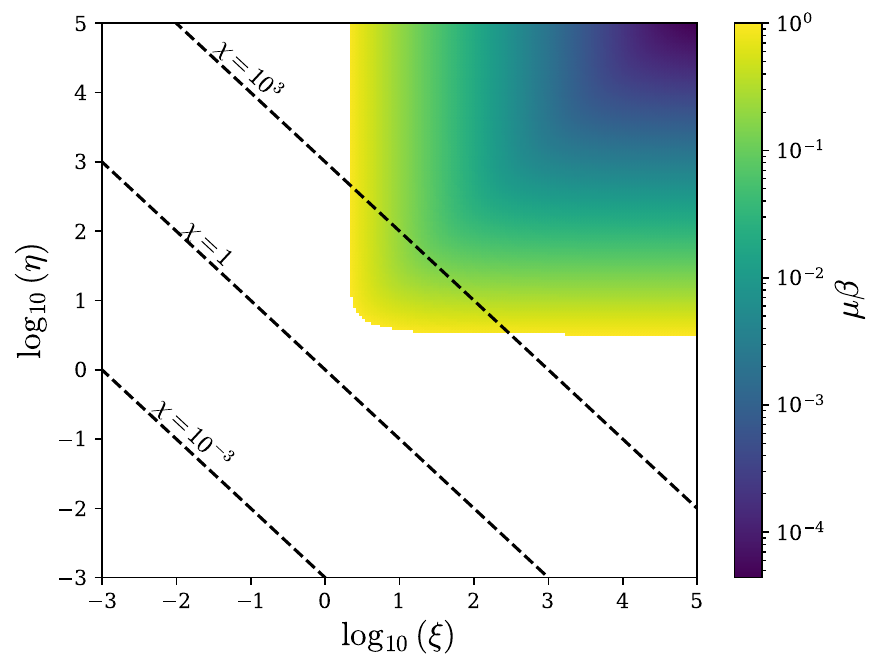}
				\caption{Length scale set by the inverse temperature $\beta$~(as in \ref{eq:thermality}) beyond which partner-mode correlations can be resolved in the strong-field/nonadiabatic regime.}
		\label{fig:PairCreationScale}
	\end{center}
\end{figure}  

Notably, the scaling-law becomes near-extensive for an ``L-shaped" region in the parameter-space. To better understand this, let us consider the length scales for which the nonlocal signatures of pair creation can be resolved. In Hawking radiation, a lower-bound for this range is set by a length scale of the order of inverse temperature $\beta_H$, which delineates a near-horizon region known as the ``quantum atmosphere" dominated by short-distance vacuum correlations~\cite{2016GiddingsPLB,2017Dey.etalPLB} --- the partner-mode correlations can only be resolved beyond this region~\cite{2021Fabbri.BalbinotPRL}. Since a similar length scale $\beta$ is available to us in the Sauter-Schwinger effect via \ref{eq:thermality}, one can draw an analogy where every point is treated as an effective horizon surrounded by a quantum atmosphere of radius $\mathcal{O}(\beta)$~(\ref{fig:PairCreationScale}). While the overall correlations proliferate in the strong-field regime, the high density of excitations generated throughout the system can also 
obscure individual pair creation signatures beyond some effective length scale $l_{\rm max}$, wherein the nonlocal correlations become \textit{noisy}. This implies that a volume-law emerges not just when there is sufficient particle content, but also when the probed length scales sufficiently resolve partner-mode correlations ($\beta\lesssim r\lesssim l_{\rm max}$). Thus even for regions where the pairwise entropy is significant, the scaling-law can drop sharply due to a narrow resolution range ($\beta \sim l_{\rm max}$), a sub-lattice range ($l_{\rm max}\ll\epsilon$) or sub-atmosphere partitions ($r\ll \beta$). Estimating $l_{\rm max}$ and isolating other potential volume-law states in this regime would however require a finer lattice as well as wider subsystem probes, and in turn, significant computing resources. Nevertheless, our analysis shows that the final state entropy scaling serves as a sensitive probe of the underlying correlation structure, recording information not just about the pairwise entanglement content, but also the spectrum of particles generated via nonperturbative dynamics.

\section{Discussion and future outlook}\label{sec:discussion}

In this work, we have carried out a detailed numerical investigation of the 
geometric scaling and dynamics of entanglement generated during nonperturbative vacuum decay in scalar QED, focusing on the Sauter-Schwinger effect. To this end, we have analyzed entanglement entropy under a cylindrical spatial bipartition and demonstrated that strong external electric fields can drive a 
fundamental change in entanglement scaling --- from the area-law behavior characteristic of vacuum states to a more general power-law scaling --- at late times. This transition provides a concrete information-theoretic signature of nonperturbative pair production and highlights entanglement entropy as a sensitive probe of strong-field quantum dynamics.

One of our key results is that this power-law scaling saturates to a volume-law behavior in appropriate strong-field regimes, particularly along an ``L-shaped" region in the $(\xi,\eta)$ parameter-space. We interpret this transition through the lens of the underlying particle spectrum, noting that the entanglement entropy in such regimes is dominated by a thermal-like distribution of low-energy pairs produced via nonperturbative dynamics. We argue that such a volume-law emerges when the probed range of subsystem sizes corresponds to the length scales where nonlocal partner-mode correlations can be sufficiently resolved. For subsystem probes beyond this range, the scaling law can drop sharply to sub-extensive values. In the intermediate regions of the pulse-profile parameter space --- where a genuine power-law is observed --- the scaling therefore indicates a competition between short-range vacuum correlations and the long-range correlations seeded by pair creation. 

Our results also clarify the relationship between momentum-space and real-space entanglement in this setting. We observe that the entropy content of the pairwise entanglement between particle–antiparticle modes --- completely characterized by the squeezing parameter and displaying the universal structure expected for quantum fields in time-dependent external backgrounds --- is reflected in the scaling pre-factor of real-space entanglement. However, the latter captures additional information about how these partner-mode correlations evolve and reorganize in space. In fact, the volume-law entanglement observed in appropriate regimes is not simply a consequence of large pairwise entropy, but instead reflects a nontrivial redistribution of correlations across many spatial modes without obscuring the nonlocal signatures of pair creation. In this way, geometric entanglement probes collective features of the final state that are not manifestly captured by purely mode-localized measures.

From a broader perspective, the observed transition in entanglement scaling raises several conceptual questions. In ground states of local Hamiltonians, area-law behavior is often linked to locality and finite correlation length, whereas volume-law scaling is usually associated with highly excited or “typical” states~\cite{1993PagePRL,2022Bianchi.etalPQ}. Our findings show that unitary, nonperturbative time evolution in a relativistic quantum field theory can dynamically generate such highly entangled states starting from the vacuum, without invoking randomness or thermalization in the usual sense. This invites a closer examination of the role of effective ergodicity, mode mixing, and  
thermal-like spectra in strong-field QFT.

There are several directions in which this work can be extended. A natural next step is to generalize the analysis to fermionic QED~\cite{2025KaushalTEPJC}, where Pauli blocking and spin degrees of freedom may qualitatively modify both pair production and entanglement scaling. It would also be interesting to explore more general pulse-profiles~\cite{2008Schuetzhold.etalPRL}, time-dependent bipartitions, or entanglement measures beyond the von Neumann entropy, such as logarithmic negativity, which can diagnose entanglement in Gaussian mixed states~\cite{2025Kranas.etal}. The methods developed here could be directly applied to other instances of particle production in curved spacetime ~\cite{1969ParkerPR,1975HawkingCiMP} and their experimental analogues~\cite{2008Balbinot.etalPRA,2004Fedichev.FischerPRA}, where entanglement plays a central role in revealing the quantum features associated with relativistic phenomena. Finally, another important direction would be to incorporate the effects of backreaction~\cite{2025Copinger.etalPRD}, and study its impact on the entanglement entropy.

\section*{Acknowledgments}
This work has been supported by the National Research Foundation of Korea under Grant No.~2020R1A2C2008103 (SMC) and STFC Consolidated Grant ST/X000494/1 “Particle Theory at the Higgs Centre” (KR).

\appendix

\section{Orthonormality and completeness of the cylindrical basis}\label{app:cylmodes_ortho}

In this section, we provide the details underlying the orthonormality and completeness of the cylindrical modes introduced in the main text. We begin by defining the positive- and negative-energy mode functions:
\begin{align}
    \varphi^{\pm}_{\kappa k_{z}m}(x)=\sqrt{2\pi}e^{ik_{z}z}e^{im\phi}J_{m}(\kappa\varrho)v^{\pm}_{\kappa k_{z}}(t)\,,
\end{align}
The Klein-Gordon inner product for two scalar fields $\varphi_1$ and $\varphi_2$ is defined by
\begin{align}
    (\varphi_1,\varphi_2)=i\int d^3\mathbfs{x}\left(\varphi^*_1\partial_t\varphi_2-\varphi_2\partial_t\varphi^*_1\right)
\end{align}
Applying this definition to the cylindrical modes, we find:
\begin{align}
    &(\varphi^{+}_{k},\varphi^{+}_{k'})=i\int d^3\mathbfs{x}\left(\varphi^*_k\partial_t\varphi_{k'}-\varphi_{k'}\partial_t\varphi^*_k\right)\nonumber\\
    &=i\left(v^*_k\dot{v}_{k'}-v_{k'}\dot{v}^*_k\right)\int_{0}^{\infty} \varrho d\varrho \int_{-\infty}^{\infty}dz\int_{0}^{2\pi} d\phi\nonumber\\
    &\qquad\times e^{i(m'-m)\phi}e^{i(k'_{z}-k_{z})z}(2\pi)J_{m}(\kappa\varrho)J_{m'}(\kappa'\varrho)\nonumber\\
    &= i\left(v^*_k\dot{v}_{k'}-v_{k'}\dot{v}^*_k\right) (2\pi)^3\delta_{m,m'}\nonumber
    \\&\qquad \qquad\qquad\times \delta(k_{z}-k_{z}')\frac{\delta(\kappa'-\kappa)}{\kappa} 
\end{align}
Here, the Wronskian $i\left(v_k^* \dot{v}_{k'} - v_{k'} \dot{v}_k^* \right)$ is independent of time. By normalizing the $v_k$ functions such that this Wronskian equals unity, we obtain the standard orthonormality relations
\begin{align}
     (\varphi^{\pm}_{k},\varphi^{\pm}_{k'})&=\pm(2\pi)^3\delta_{m,m'}\delta(k_{z}-k_{z}')\frac{\delta(\kappa'-\kappa)}{\kappa}\\
     (\varphi^{\pm}_{k},\varphi^{\mp}_{k'})&=0
\end{align}
To verify completeness, one considers the matrix
\begin{align}
    &-i\left[\varphi^{+}_{k}(t,\mathbf{x})\partial_{t}\varphi^{+*}_{k}(t,\mathbf{x}')-\varphi^{-}_{k}(t,\mathbf{x})\partial_{t}\varphi_{k}^{-*}(t,\mathbf{x}')\right]\nonumber\\
    &= e^{i(\phi-\phi')m}\,e^{ik_{z}(z-z')}(2\pi)J_{m}(\kappa\varrho)J_{m}(\kappa\varrho')
\end{align}
and summing over all modes gives
\begin{align}
  &\sum_{k}-i\left[\varphi^{+}_{k}(t,\mathbf{x})\partial_{t}\varphi^{+*}_{k}(t,\mathbf{x}')-\varphi^{-}_{k}(t,\mathbf{x})\partial_{t}\varphi_{k}^{-*}(t,\mathbf{x}')\right]\nonumber\\
  &= \sum_{m=-\infty}^{\infty}\int_{-\infty}^{\infty} \frac{dk_{z}}{(2\pi)}\int \frac{\kappa d\kappa}{(2\pi)^2} e^{i(\phi-\phi')m}\,e^{ik_{z}(z-z')}\nonumber\\&\qquad\qquad\qquad\qquad\qquad\times(2\pi)J_{m}(\kappa\varrho)J_{m}(\kappa\varrho')\nonumber\\
  &=\delta(\phi-\phi')\delta(z-z')\frac{\delta(\varrho-\varrho')}{\varrho},
\end{align}
confirming the completeness of the cylindrical basis.\\

\subsection{Details of discrete modes}

We now impose boundary conditions leading to discretized modes. Starting from the continuum expressions
\begin{align}
    \varphi^{\pm}_{\kappa k_{z}m}(x)=\sqrt{2\pi}e^{ik_{z}z}e^{im\phi}J_{m}(\kappa\varrho)v^{\pm}_{\kappa k_{z}}(t)\,,
\end{align}
and imposing periodic condition along the axis, we get
\begin{align}
    k_{z}L=2\pi l\Rightarrow k_{z,l}=\frac{2\pi l}{L} \,,
\end{align}
so that the modes become
\begin{align}
    \varphi^{\pm}_{\kappa k_{z}m}(x)&\rightarrow \varphi^{\pm}_{\kappa l m}(x)\\&=\sqrt{2\pi}e^{i\frac{2\pi l}{L}z}e^{im\phi}J_{m}(\kappa\varrho)v^{\pm}_{\kappa k_{z}}(t)\,.\nonumber
\end{align}
Similarly, imposing a Dirichlet condition at the radial boundary $\varrho=R$ leads to
\begin{align}
   J_{m}(\kappa R)=0\Rightarrow \kappa_{mn}=\frac{\jj_{mn}}{R} \,,
\end{align}
where $\jj_{mn}$ denotes the $n^{\rm th}$ zero of $J_m$. The normalized discrete modes then read
\begin{align}
    \varphi^{\pm}_{\kappa k_{z}m}(x)&\rightarrow \varphi^{\pm}_{n l m}(x)\\&=\frac{2\sqrt{\pi}e^{i\frac{2\pi l}{L}z}e^{im\phi}}{|J_{m+1}(\jj_{mn})|}J_{m}\left(\frac{\jj_{mn}\varrho}{R}\right)v^{\pm}_{\kappa k_{z}}(t)\,.\nonumber
\end{align}
The corresponding orthonormality relations for these discrete modes are
\begin{align}
     (\varphi^{\pm}_{k},\varphi^{\pm}_{k'})&=\pm\mathcal{V}\delta_{m,m'}\delta_{l,l'}\delta_{n,n'}\,,\\
     (\varphi^{\pm}_{k},\varphi^{\mp}_{k'})&=0\,,
\end{align}
where $\mathcal{V}=\pi R^2L$ the total volume. This motivates the discrete mode sum prescription
\begin{align}
    \sum_{k}=\frac{1}{\mathcal{V}}\sum_{m=-\infty}^{\infty}\sum_{l=-\infty}^{\infty}\sum_{n=0}^{\infty}
\end{align}
which will be used in computations of mode sums, entanglement entropy, and related observables.

\bibliography{EntQED}

\end{document}